\documentclass[preprint,12pt,authoryear]{elsarticle}

\usepackage{amssymb}
\usepackage{lmodern} 
\usepackage{graphicx} 
\usepackage{amsmath}
\usepackage{amsthm}
\usepackage{amsfonts}
\usepackage{bbm}
\usepackage{float}
\usepackage {wrapfig} 
\usepackage{subcaption}
\usepackage{multicol}
\usepackage{multirow}
\usepackage{enumerate}
\usepackage{textcomp}
\newtheorem{theorem}{Theorem}
\newtheorem{remark}[theorem]{Remark}
\usepackage[hidelinks]{hyperref}
\usepackage[top = 2.5cm, bottom = 2.5cm, left = 2.5cm, right = 2.5cm]{geometry}

\journal{IJEPES}

\begin{document}

\begin{frontmatter}



\title{Bootstrap prediction regions for daily curves of electricity demand and price using functional data}


\author[inst1]{Rebeca Pel\'{a}ez}
\author[inst1]{Germ\'{a}n Aneiros}
\author[inst1]{Juan Vilar}

\affiliation[inst1]{organization={Research Group MODES, Department of Mathematics, Faculty of Computer Science, Universidade da Coruña},
	city={A Coruña},
	postcode={15071}, 
	country={Spain}}

\begin{abstract}
The aim of this paper is to compute one-day-ahead prediction regions for daily curves of electricity demand and price. Three model-based procedures to construct general prediction regions are proposed, all of them using bootstrap algorithms. The first proposed method considers any $L_p$ norm for functional data to measure the distance between curves, the second one is designed to take different variabilities along the curve into account, and the third one takes advantage of the notion of depth of a functional data. The regression model with functional response on which our proposed prediction regions are based is rather general: it allows to include both endogenous and exogenous functional variables, as well as exogenous scalar variables; in addition, the effect of such variables on the response one is modeled in a parametric, nonparametric or semi-parametric way. A comparative study is carried out to analyse the performance of these prediction regions for the electricity market of mainland Spain, in year 2012. This work extends and complements the methods and results in \cite{Aneiros2016} (focused on curve prediction) and \cite{Vilar2018} (focused on prediction intervals), which use the same database as here.
\end{abstract}


\end{frontmatter}



\section{Introduction}
\label{sec:Introduction}

Prediction is one of the main aims of time series analysis. In fact, having at hand accurate future values of some variable of interest enables managers or policy makers to make properly informed decisions. In the case of electricity markets, given that electricity cannot be stored, it is very important for producers to forecast future values of electricity demand to anticipate to future demands and to avoid overproduction. In the same way, knowing proper future values of electricity price allows agents and companies involved in the electricity market to design appropriate strategies, contributing to anticipate scenarios to make timely decisions.

In the last decades, a lot of research was focused on electricity demand and price forecasting. Most of such research considered the case where all the involved variables are scalar (real-valued variables), focusing mainly on hourly and daily demand (or price) forecasting at different forecast horizons (one day ahead for hourly forecasts and several days ahead for daily forecasts). For time-series methodology related to electricity demand and price forecasting, see for instance the reviews by \cite{Suganthi2012}, \cite{Weron2014}, \cite{Hong2016} and \cite{Nowotarski2018}, as well as \cite{Weron2006} for a monograph. See also \cite{Kaya2023} for a recent study based on deep learning approaches (artificial neural networks).

Nowadays, with the development of modern technology together with the high storage capacity, it is usual to deal with functional data (function-valued variables); that is, data observed on a continuum, often time but not only. Some examples of functional data are curves (daily electricity demand or price curves, \ldots), images (magnetic resonance imaging, \ldots), etc. From the 1990s, Functional Data Analysis (FDA) has emerged as one of the most challenging tasks in Statistics, and many inference techniques (estimation, testing of hypotheses, variance analysis, prediction, classification, etc) were extended from multidimensional data (finite dimensional) to functional data (infinite dimensional). See, for instance, \cite{Ramsay2005} and \cite{Ferraty2006} for some early monographs on parametric and nonparametric modelling of functional data, respectively, and the more recent monographs by \cite{Zhang2014} and \cite{Kokoszka2017} for variance analysis and an introduction to several topics in FDA, respectively; see also \cite{Aneiros2022} for a compendium of recent advances in FDA.

The case of electricity demand and price forecasting from models including functional data was also considered in the literature (which is much more limited than that of the scalar case). A pioneer work was that of \cite{Vilar2012}, where hourly electricity demand and price were predicted from both functional nonparametric and semi-functional partially linear regression models with scalar response (hourly electricity demand or price) and some functional explanatory variable (lagged daily curves of electricicity demand or price); in the case of the semi-functional partially linear model, dummy variables were also included to indicate the type of day. See \citealt{Ferraty2004} and \citealt{Aneiros2008} for some first theoretical results on the two referred regression models, respectively. The article  by \cite{Vilar2012} was extended in \cite{Ranha2018}, where additive models were considered; in addition, some exogenous scalar covariates were included in the models. \cite{Liebl2013} proposed a new statistical perspective for modelling and forecasting electricity spot prices that accounts for the merit order model, where a functional factor model is used to parameterise the series of daily price-demand functions. Recently, \cite{Lisi2020} investigated the forecasting performance of several models for the 1-day-ahead prediction of demand and price on four electricity markets: nonparametric procedures are used to estimate the deterministic component, while the functional approach takes part in the modelling of the residual stochastic component.

All the articles referred to in the previous paragraph are focused on prediction of a scalar response variable (in general, hourly or daily electricity demand or price) using information from some functional explanatory variable. \cite{Antoch2010} was one of the first papers devoted to prediction of curves of electricity consumption. More specifically, they predicted the forthcoming weekend --and weekdays-- consumption curves from functional linear regression models, where both the response and explanatory variables were of functional nature. \cite{Aneiros2016} extended the procedures in \cite{Vilar2012} in two ways: on the one hand, some exogenous scalar covariates were added to the models (temperature when the aim was demand prediction, and forecasted daily demand and wind power production for the case of price prediction); on the other hand, the response variable was functional: daily electricity demand and price curves. Among the conclusions obtained in \cite{Aneiros2016}, it is worth highlighting: 
`to take information from exogenous covariates improves the forecasts, hourly predictions obtained from the discretisacion of the predicted curves are better than the predictions given by the corresponding functional models with scalar response, and the computational time to obtain such hourly predictions from discretisation is much lower than the needed time to implement the corresponding scalar procedures'. Some other related works dealing with functional response (more specifically, focused on prediction of interest curves in the electric power market) are \cite{Aneiros2013}, \cite{Paparoditis2013}, \cite{Chen2017}, \cite{Portela2018}, \cite{Elias2022} and \cite{Barrientos2023}.

This paper is motivated by the conclusions in \cite{Aneiros2016} referred in the previous paragraph. The aim of this paper is twice: on the one hand, to propose methodology to construct prediction regions for a future curve in a setting of functional time series; on the other hand, to put in practice such regions for the case of daily electricity demand and price curves. A prediction region, at a predetermined confidence level $1-\alpha$, for a random variable $\zeta_{N+1}$ is a random region in which $\zeta_{N+1}$ will take values with probability $1-\alpha$. In the case that $\zeta_{N+1}$ is real valued, the term `prediction interval' is used instead of `prediction region'. At this moment, it is worth being noted the main difference between prediction intervals and prediction regions. To clarify this, let us consider the example we will discuss in Section \ref{sec:Application}, where a curve to predict will be the daily curve of electricity demand, say $\{\zeta_{N+1}(t); t \in (0,24]\}$. If $R_{\alpha}=\{(LR_{N+1}(t),UR_{N+1}(t)); t \in (0,24]\}$ is a prediction region for $\zeta_{N+1}$ at confidence level $1-\alpha$, then it verifies that
\begin{equation*} 
	P\left(LR_{N+1}(t) < \zeta_{N+1}(t) < UR_{N+1}(t); \forall t \in (0,24]\right)=1-\alpha. 
\end{equation*}
Nevertheless, if one computes a prediction interval for $\zeta_{N+1}(t)$, say $(LI_{N+1}(t),UI_{N+1}(t))$, for each $t \in (0,24]$, what is verified is that
\begin{equation*} 
	P\left(LI_{N+1}(t) < \zeta_{N+1}(t) < UI_{N+1}(t)\right)=1-\alpha, \ \forall t \in (0,24],
\end{equation*}
but not
\begin{equation*} 
	P\left(LI_{N+1}(t) < \zeta_{N+1}(t) < UI_{N+1}(t); \forall t \in (0,24]\right)=1-\alpha.
\end{equation*}
In particular, when one computes both $R_{\alpha}$ and $I_{\alpha}=\{(LI_{N+1}(t),UI_{N+1}(t)); t \in (0,24]\}$ from observed data, one should expect that the area of $R_{\alpha}$ is greater than the one of $I_{\alpha}$.

The topic of prediction intervals for electricity demand and price using functional data was dealt, for instance, in \cite{Vilar2018} and \cite{Ranha2018}. In particular, they constructed model-based prediction intervals using bootstrap procedures, and such intervals were applied to hourly electricity demand and price. The case of prediction regions in a context of functional time series was studied, from a theoretical point of view, in \cite{Zhu2017}; as in the case of \cite{Vilar2018} and \cite{Ranha2018} for prediction intervals, the prediction regions proposed in \cite{Zhu2017} were based on both a regression model and bootstrap techniques. In this paper, we propose three model-based approaches to construct prediction regions using bootstrap algorithms: the first approach extends the proposal in \cite{Zhu2017} from a functional autoregressive nonparametric model to a fairly general functional model (allowing to include both endogenous and exogenous functional variables, as well as exogenous scalar variables); the second approach is designed to take different variabilities along the curve into account; finally, the third approach is based on the notion of depth of a functional data. Then, these three approaches will be used to compute and compare prediction regions for future daily curves of electricity demand and price in the Spanish Electricity Market. 

The rest of this work is organised as follows. The three model-based proposals to construct prediction regions are presented in Section \ref{sec:Prediction_regions}. More specifically: our prediction regions depends on a fairly general regression model, which is stated in Section \ref{general_setting}; then, Section \ref{sec:Prediction_regions_building} is devoted to motivate both the construction of each prediction region and the corresponding bootstrap algorithms; because of the regression function is unknown, Section \ref{sec:Prediction_regions_forecasting} presents two particular regression functions and their corresponding estimators; some proposals to choose the tuning parameters related to our procedures are given in Section \ref{tuning}. Our methodology to construct prediction regions is applied to data related to the electricity market of mainland Spain, in years 2011 and 2012, and they are presented in Section \ref{sec:Data}. The obtained results are reported and compared in Section \ref{sec:Application}. Finally, Section \ref{sec:Conclusions} provides some conclusions.

\section{Prediction regions for functional data}
\label{sec:Prediction_regions}

This section contains the main methodological contribution of this paper: three proposals to construct prediction regions for functional time series. Such proposals are both model based and bootstrap based. Firstly we establish the considered general model and then we construct the proposed bootstrap prediction regions. Some particular models and a guide to select the tuning parameters are also included.

\subsection{General setting}
\label{general_setting}
Let $\left\{ \zeta \left(t\right) \right\} _{t\in R}$ be a real valued continuous time seasonal stochastic process with seasonal length $\tau$. Then, assuming that such process is observed on the interval $I=(a,b] \subset R$ with $b = a + N\tau$, one can cut $\left\{ \zeta \left(t\right) \right\} _{t\in I}$ into $N$ succesive random curves, $\left\{ \zeta _{i}\right\} _{i=1}^{N}$, where 
\begin{equation*}
	\zeta _{i}\left( t\right) =\zeta \left( a+\left( i-1\right) \tau +t\right),%
	\text{with} \ t \in \left( 0,\tau \right]. 
\end{equation*}
And finally, it seems natural to take advantage of the dependence between the random curves, $\zeta _{i}$, in order to predict $\zeta _{N+1}$ using information given from the discrete time stochastic process, $\left\{ \zeta _{i}\right\} _{i=1}^{N}$. Note that, mathematically, $\zeta _{i}$ is a functional data that take values in the space of real functions defined 
in $\left( 0,\tau \right]$. Such space will be denoted by $\mathcal{F}$. It is a separable Hilbert space, so it is endowed with an inner product $\left\langle \cdot ,\cdot \right\rangle $ and the corresponding norm, $\left\Vert \cdot \right\Vert$, defined by  $\left\Vert
\zeta \right\Vert ^{2}=\left\langle \zeta ,\zeta \right\rangle$. 

In addition to $\zeta _{i}$ ($i=1,\ldots,N$), some exogenous variables (say $\mathbf{x}_{i}$ and $\mathbf{\xi }_{i}$) could be relevant in predicting $\zeta _{N+1}$. In this paper, we will obtain model-based prediction regions. More specifically, we will assume that the following regression model holds: 
\begin{equation}
	\label{general_model}
	\zeta _{i}=r(\zeta _{i-1},\mathbf{x}_{i},
	\mathbf{\xi }_{i})+\varepsilon _{i}, \quad i \in S \subset \mathcal{I}=\{2,\ldots,N\},
\end{equation}
where $\mathbf{x}_{i}$ and $\mathbf{\xi }_{i}$ are $p$-dimensional and $q$-dimensional vectors of exogenous scalar and functional variables, respectively. The unknown operator $r(\cdot)$ denotes the regression function while $\{\varepsilon _{i}\}$ are the random functional errors with zero mean and independent and identically distributed. From now on, $n_S$ will denote the cardinal of $S$, while the vector of explanatory variables in model (\ref{general_model}), $(\zeta _{i-1}, \mathbf{x}_{i},\mathbf{\xi }_{i})$, will be denoted by $\chi_{i}$.

Note that model (\ref{general_model}) is a fairly general model, allowing to include the effect of an endogenous variable and exogenous variables of scalar and/or functional nature. Note also that, in order to make clearer our proposal, we have considered only one endogenous variable (autoregression of order one), although the procedure could be generalised to higher orders.


\subsection{Building the prediction regions}
\label{sec:Prediction_regions_building}

A prediction region for $\zeta _{N+1}$ at confidence level $1 - \alpha$ is a random subset $R_{\alpha}\subseteq \mathcal{F}$, such that
\begin{equation*}
	P(\zeta_{N+1}\in R_{\alpha}) = 1-\alpha.
\end{equation*}
From this definition, and taking into account 
model (\ref{general_model}), it follows that our prediction region depends on 
the operator $r(\cdot)$, which is unknown. From now on, $\widehat{r}_h(\cdot)$ will denote a nonparametric estimator of $r(\cdot)$, with $h$ being a smoothing parameter. Section \ref{sec:Prediction_regions_forecasting} will present the regression functions, $r(\cdot)$, and the corresponding estimators, $\widehat{r}_h(\cdot)$, explored in this study.

In this paper, three methods for computing prediction regions for $\zeta _{N+1}$ from information given by the sample $\left\{ \left( \chi _{i},\zeta_{i}\right), i \in S \right\}$ are proposed. The three methods are based on an algorithm for bootstrap resampling of the such sample. The algorithm is shown below:




\paragraph{Algorithm for Bootstrap Resampling (BR)} 


\begin{enumerate}
	
	\item Compute $\widehat{r}_{b}(\chi _{i}), \ i\in S$, that is, the prediction of $\zeta _{i}|\chi _{i}$, where $b>0$ is a smoothing parameter.
	
	\item Compute the centered residuals,
	\begin{equation}
		\widehat{\epsilon}_{i,b}=\widehat{%
			\varepsilon }_{i,b}-\widehat{\bar{\varepsilon}}_{b}, \ i\in S, \label{res}
	\end{equation}
	where $\widehat{\varepsilon }_{i,b}=\zeta
	_{i}-\widehat{r}_{b}\left( \chi _{i}\right)$ and $\widehat{\bar{
			\varepsilon}}_{b}=n_{S}^{-1}\sum_{i\in S}\widehat{\varepsilon }_{i,b}$.
	
	\item Draw $n_S$ i.i.d. random variables from the empirical distribution of $\{\widehat{\epsilon}_{i,b}\}_{i \in S}$. Such variables are denoted by $\widehat{\varepsilon}%
	_{i}^{\ast }, \ i \in S$ (bootstrap pseudo-residuals). 
	
	\item Compute
	$\zeta _{i}^{\ast }=\widehat{r}_{b}(\chi _{i})+\widehat{\varepsilon }%
	_{i}^{\ast }, \ i \in S$.
	
	\item  Obtain  $\left\{ \left( \chi _{i},\zeta 	_{i}^{\ast }\right), i \in S \right\}$, a bootstrap resample of the original sample $\left\{ \left( \chi _{i},\zeta_{i}\right), i \in S \right\}$.
	
\end{enumerate}

The following Sections \ref{Lp-method}, \ref{lambda-method} and \ref{depth-method} present our three proposed methods to construct prediction regions.

\subsubsection{$L_p$-method for prediction regions}
\label{Lp-method}

In this section, we assume that the observed curve $\zeta _{N+1}|\chi _{N+1}$ belongs to the normed functional space $L_p(\mathcal{F})\equiv(\mathcal{F},\|\cdot\|_p)$, where the norm $\|\cdot\|_p$ is defined using a natural generalisation of the $p$-norm for finite-dimensional vector spaces.

The paper by \cite{Zhu2017} was one of the first articles in the statistical literature where model-based prediction regions were proposed in a setting of functional time series. Such prediction regions were based on functional autorregression models of order 1. The method we propose here extends the proposal in \cite{Zhu2017} to the case of the general model in \eqref{general_model}.

If we consider a value $\rho_{\alpha}>0$ verifying
$$
P\big(  \big\|\zeta _{N+1}|\chi _{N+1} - \widehat{r}_h(\chi _{N+1})  \big\|_{p} < \rho_{\alpha}  \big) = 1 - \alpha,
$$
then,
$$ R_{\rho_\alpha}  =\big\{ \varphi \in L_p(\mathcal{F}): \; \big\| \varphi -   \widehat{r}_h(\chi_{N+1}) \big\|_{p} < \rho_{\alpha} \big\}
$$
is a prediction region for $\zeta _{N+1}|\chi _{N+1}$ computed at the $1 - \alpha$ confidence level (note that $R_{\rho_\alpha}$ is the ball around $\widehat{r}_h(\chi _{N+1}) $ of radius $\rho_{\alpha}$, which is denoted by $B( \widehat{r}_h(\chi _{N+1}) , \rho_{\alpha} ))$. But, in practice, $R_{\rho_\alpha}$ is infeasible because $\rho_\alpha$ depends on the distribution of $\|\zeta _{N+1}|\chi _{N+1} - \widehat{r}_h(\chi _{N+1})\|_{p}$, which is unknown. 

We propose to approximate the distribution of $\|\zeta _{N+1}|\chi _{N+1} - \widehat{r}_h(\chi _{N+1})\|_{p}$ by means of a bootstrap procedure, which takes adventage of the equality (see (\ref{general_model}))
\begin{equation}
	\zeta _{N+1}|\chi _{N+1} - \widehat{r}_h(\chi _{N+1})  = 
	r(\chi _{N+1}) -\widehat{r}_h( \chi _{N+1})  
	+ \varepsilon _{N+1}|\chi _{N+1}. \label{descomp}
\end{equation}
In the following, a bootstrap approximation for $\| r(\chi _{N+1}) -\widehat{r}_h( \chi _{N+1}) 
+ \varepsilon _{N+1}|\chi _{N+1}\|_p$ is given by $ \| \widehat{r}_b(\chi_{N+1})  - \widehat{r}_{h}^*(\chi_{N+1}) + \varepsilon^{*}_{N+1} \|_{p} $. 
Then, our proposed bootstrap prediction region is $R_{\rho_\alpha^*}$, 
where  $\rho_{\alpha}^*$ is such that
$$
P^*\big(   \big\| \widehat{r}_b(\chi_{N+1})   - \widehat{r}_{h}^*(\chi_{N+1}) + \varepsilon^{*}_{N+1} \big\|_{p} < \rho_{\alpha}^*    \big) = 1 - \alpha.
$$

The algorithm to obtain the prediction region for $\zeta_{N+1}$ at confidence level $1 - \alpha$ is explained below. The Monte Carlo method is used to approximate the radius $\rho_{\alpha}^*$, so that the prediction region, $R_{\rho_\alpha^*}$, has a confidence level approximately equal to $1-\alpha$.

\paragraph{Algorithm of the $L_p$ method for prediction regions}


\begin{enumerate}
	
	\item Compute the predictor for $\zeta _{N+1}$ given by $\widehat{\zeta }_{N+1} = \widehat{r}_{h}(\chi _{N+1})$ with bandwidth $h$. 
	
	\item Using the algorithm BR, obtain the bootstrap resample
	$\zeta _{i}^{\ast }=\widehat{r}_{b}(\chi _{i})+\widehat{\varepsilon }%
	_{i}^{\ast }, \ i \in S$. (Note that $\zeta_{i}^{\ast }$ depends on $b$; for the sake of clarity, we have written $\zeta_{i}^{\ast}=\zeta_{b,i}^{\ast}$.)
	
	\item Compute the bootstrap predictor of $\zeta_{N+1}^*$ from the bootstrap resample $\left\{ \left( \chi _{i},\zeta
	_{i}^{\ast }\right), i\in S\right\}$ as follows:
	\begin{equation*}
		\widehat{\zeta}_{N+1}^{\ast }=\widehat{r}_{h}^{\ast }\left( \chi _{N+1}\right).
	\end{equation*}
	
	\item Repeating $B$ times Steps 2-3, obtain the $B$ bootstrap
	predictors 
	$$\left\{ \widehat{\zeta}_{N+1}^{\ast ,j}\right\} _{j=1}^{B} = \left\{ 
	\widehat{r}_{h}^{\ast ,j}\left( \chi _{N+1}\right) \right\} _{j=1}^{B}.$$
	
	\item Draw $B$ i.i.d random variables $\left\{ \varepsilon
	_{N+1}^{\ast, j }\right\} _{j=1}^{B}$ from the empirical distribution
	of the centered residuals $\{\widehat{\epsilon}_{i,b}\}_{i \in S}$ (see (\ref{res})).
	
	\item For $j = 1, \dots, B$, obtain the future bootstrap observation 
	\begin{equation*}
		\zeta _{N+1}^{\ast, j}=\widehat{r}_{b}(\chi _{N+1})+\varepsilon _{N+1}^{\ast, j }.
	\end{equation*}
	
	\item For $j = 1, \dots, B$, compute the bootstrap error: 
	\begin{equation*}
		E_j^* = 	\zeta _{N+1}^{\ast, j}  -  \widehat{\zeta}_{N+1}^{\ast ,j}   =  \widehat{r}_{b}(\chi _{N+1})-\widehat{r}_{h}^{\ast ,j}\left(
		\chi _{N+1}\right) +\varepsilon _{N+1}^{\ast, j }.
	\end{equation*}
	
	\item For $j = 1, \dots, B$, obtain
	\begin{equation*}
		\rho_{j}^{\ast }= \|  E_j^*  \|_p. 
	\end{equation*}
	
	\item Sort the values $\rho^*_{1}, \dots, \rho^*_{B}$, obtaining $\rho^*_{(1)}, \dots, \rho^*_{(B)}$ and select $\rho^*_{\alpha} = \rho^*_{([B(1 - \alpha)])}$.
	
	\item The $\left( 1-\alpha \right) 100\%$ prediction region for $\zeta _{N+1}|\chi _{N+1}$ consists of all $\varphi$
	such that 
	\begin{equation*}
		\| \varphi -\widehat{r}_h\left( \chi _{N+1}\right) \|_p \leq
		\rho^*_{\alpha}.
	\end{equation*}
	
\end{enumerate}

\begin{remark}
	\label{remark1}
	It is worth being noted that, given the popularity of the $L_p$ norms, in this section we have considered that kind of norms; so the method was named $L_p$-method. Actually, everything presented in this section is valid if instead of the norm $\|\cdot\|_p$ one uses any other norm in the functional space $\mathcal{F}$.
\end{remark}

\subsubsection{$\lambda$-method for prediction regions}
\label{lambda-method}


The $\lambda$-method for prediction regions is based on finding the value of $\lambda_{\alpha} >0$ 
such that 
$$
P\big(  \big|\zeta _{N+1}(t)|\chi _{N+1} - \widehat{r}_h(\chi _{N+1})(t)  \big| < \lambda_{\alpha}  \sigma(t), \; \forall t \in (0,\tau] \big)
= 1 - \alpha,
$$
where $\sigma^2(t) = Var \big( \widehat{r}_h(\chi _{N+1})(t) \big) \ \forall t \in (0,\tau]$. Thus, the theoretical prediction region at confidence level $1-\alpha$ is defined by
$$ 
R_{\lambda_{\alpha}, \sigma} = \Big\{ \varphi \in \mathcal{F}: \varphi(t) \in \big(  \widehat{r}_h(\chi _{N+1})(t)  - \lambda_{\alpha}\sigma(t),    \widehat{r}_h(\chi _{N+1})(t)  + \lambda_{\alpha}\sigma(t)  \big)      , \; \forall t \in (0,\tau]  \Big\}.
$$
As in the case of our first proposed prediction region, $R_{\rho_\alpha}$ (see Section \ref{Lp-method}), $R_{\lambda_{\alpha}, \sigma}$ is infeasible: $\lambda_{\alpha}$ depends on both the unknown parameter $\sigma(t)$  and the unknown distribution of $|\zeta _{N+1}(t)|\chi _{N+1} - \widehat{r}_h(\chi _{N+1})(t)  \big|$. We propose a bootstrap algorithm to approximate $\lambda_{\alpha}$ and $\sigma(t)$ by means of $\lambda_{\alpha}^*$ and $\sigma^*(t)$. More specifically, according to \eqref{descomp}, the values of $\lambda_{\alpha}^*$ and $\sigma^*(t)$ must satisfy that
\begin{equation}
	\label{eq:Bootstrap_p_lambda}
	p(\lambda_{\alpha}^*) = P^*\big(   \big| \widehat{r}_b(\chi_{N+1})(t)   - \widehat{r}_{h}^*(\chi_{N+1})(t) + \varepsilon^{*}_{N+1}(t) \big| < \lambda_{\alpha}^* \sigma^* (t)   , \; \forall t \in (0,\tau]  \big) = 1 - \alpha,
\end{equation}
where 
$b$ is some auxiliary smoothing parameter.

The algorithm to obtain the bootstrap prediction region for $\zeta _{N+1}|\chi _{N+1}$ at confidence level $1 - \alpha$ is explained below. The Monte Carlo method is used to approximate $\sigma^*(t)$, and an iterative method is used to approximate the value of $\lambda_{\alpha}^*$ so that the proposed prediction region, $R_{\lambda_{\alpha}^*, \sigma^*}$, has a confidence level approximately equal to $1-\alpha$.

\paragraph{Algorithm of the $\lambda$-method for prediction regions}


\begin{enumerate}
	
	\item Compute the predictor for $\zeta _{N+1}$ given by $\widehat{\zeta }_{N+1} = \widehat{r}_{h}(\chi _{N+1})$ with bandwidth $h$.
	
	\item Using the algorithm BR, obtain the bootstrap resample
	$\zeta _{i}^{\ast }=\widehat{r}_{b}(\chi _{i})+\widehat{\varepsilon }%
	_{i}^{\ast }$, with $i \in S$.
	
	\item Compute the bootstrap predictor of $\zeta_{N+1}^*$ from the bootstrap resample $\left\{ \left( \chi _{i},\zeta
	_{i}^{\ast }\right), i\in S \right\}$ as follows:
	\begin{equation*}
		\widehat{\zeta}_{N+1}^{\ast }=\widehat{r}_{h}^{\ast }\left( \chi _{N+1}\right).
	\end{equation*}
	
	\item Repeating $B$ times Steps 2-3, obtain the $B$ bootstrap
	predictors 
	$$\left\{ \widehat{\zeta}_{N+1}^{\ast ,j}\right\} _{j=1}^{B} = \left\{ 
	\widehat{r}_{h}^{\ast ,j}\left( \chi _{N+1}\right) \right\} _{j=1}^{B}.$$
	
	\item Draw $B$ i.i.d random variables $\{ \varepsilon
	_{N+1}^{\ast, j }\} _{j=1}^{B}$ from the empirical distribution
	of the centered residuals $\{\widehat{\epsilon}_{i,b}\}_{i \in S}$ (see (\ref{res})).
	
	\item Approximate the standard deviation of $\widehat{r}_h(\chi_{N+1})(t)  $ by 
	$$
	{\sigma}^*(t) \simeq  \Bigg( \dfrac{1}{B} \sum_{j = 1}^{B}  \bigg( \widehat{r}_{h}^{\ast ,j}(\chi _{N+1})(t) -\dfrac{1}{B} \sum_{k = 1}^{B} \widehat{r}_{h}^{\ast ,k}(\chi _{N+1})(t) \bigg)^2   \Bigg)^{1/2}.
	$$
	\item Use an iterative method to obtain an approximation of the value $\lambda_{\alpha}^*$ defined in \eqref{eq:Bootstrap_p_lambda}.
	
	\item The $\left( 1-\alpha \right) 100\%$ prediction region for $\zeta _{N+1}|\chi _{N+1}$ consists of all $\varphi$
	such that 
	\begin{equation*}
		| \varphi(t) -\widehat{r}_h\left( \chi _{N+1}\right)(t) | \leq
		{\lambda}_{\alpha}^*{\sigma}^*(t),
	\end{equation*}
	for all $ t \in (0,\tau]$.

\end{enumerate}

\paragraph{Iterative method to approximate $\mathbf{\boldsymbol{\lambda}_{\boldsymbol{\alpha}}^*}$}

The iterative method to approximate the value of $\lambda_{\alpha}^*>0$ 
so that the bootstrap prediction region has a confidence level approximately equal to $1-\alpha$ is explained below. This algorithm, which follows ideas of \cite{Cao2010}, allows to quickly and efficiently approximate the parameter $\lambda_{\alpha}^*$.

Let $\left\{ \widehat{\zeta}_{N+1}^{\ast ,j}\right\} _{j=1}^{B} = \left\{ 
\widehat{r}_{hb}^{\ast ,j}(\chi _{N+1}) \right\} _{j=1}^{B}$ be the bootstrap predictors of $\zeta _{N+1}$ and $\{ \varepsilon
_{N+1}^{\ast, j }\} _{j=1}^{B}$ the bootstrap model errors. Define the Monte Carlo approximation of $p(\lambda)$ in \eqref{eq:Bootstrap_p_lambda}, for any $\lambda >0$, 
as follows:
\begin{equation}
	\label{eq:Bootstrap_p_lambda_MonteCarlo}
	{p}(\lambda) \simeq \dfrac{1}{B} \sum_{j = 1}^B 
	I \Big(  
	\big| \widehat{r}_b(\chi_{N+1})(t)  - \widehat{r}_{h}^{\ast ,j}(\chi_{N+1})(t) + \varepsilon^{*, j}_{N+1} (t)  \big| < \lambda \sigma^*(t) , \; \forall t \in (0,\tau]
	\Big).
\end{equation}
Let $\lambda_L, \lambda_H >0$ 
be such that 
$  p(\lambda_L) \leq 1 - \alpha \leq p(\lambda_H)$
and let $\eta > 0$ be a tolerance, for example, $\eta = 10^{-4}$.
\begin{enumerate}
	\item Obtain $\lambda_M = \dfrac{\lambda_L + \lambda_H}{2}$ and compute Monte Carlo approximations of  ${p}(\lambda_L)$, ${p}(\lambda_M)$ and ${p}(\lambda_H)$ according to \eqref{eq:Bootstrap_p_lambda_MonteCarlo}. 
	
	\item  If $p(\lambda_M) = 1- \alpha$ or $p(\lambda_H) - p(\lambda_L) < \eta$, then  $\lambda_{\alpha}^{*} = \lambda_M$. Otherwise,
	
	\begin{enumerate}
		\item If $1 - \alpha < {p}(\lambda_M)$, then $\lambda_H = \lambda_M$ and return to Step 1.
		\item If ${p}(\lambda_M) < 1 - \alpha $, then $\lambda_L = \lambda_M$ and return to Step 1.
	\end{enumerate}	
\end{enumerate}

\subsubsection{Depth-based method for prediction regions}
\label{depth-method}

\cite{Elias2022} proposed depth-based methods for functional time series forecasting. In this section, some of such ideas are used to construct prediction regions. In any case the method proposed in this paper differs from the one proposed in \cite{Elias2022} since ours involves bootstrap procedures in the computation of prediction regions. The proposed algorithm is defined below.

\paragraph{Algorithm of the depth-based method for prediction regions} 

\begin{enumerate}
	
	\item Using the algorithm BR, obtain the bootstrap resample
	$\zeta _{i}^{\ast }=\widehat{r}_{b}(\chi _{i})+\widehat{\varepsilon }_{i}^{\ast }$, with $i \in S$.
	
	\item Compute the bootstrap predictor of $\zeta_{N+1}^*$ from the bootstrap resample $\left\{ \left( \chi _{i},\zeta_{i}^{\ast }\right), i\in S\right\}$ as follows:
	\begin{equation*}
		\widehat{\zeta}_{N+1}^{\ast }=\widehat{r}_{h}^{\ast }\left( \chi _{N+1}\right).
	\end{equation*}
	
	\item Repeating $B$ times Steps 1-2, obtain the $B$ bootstrap
	predictors 
	$$\left\{ \widehat{\zeta}_{N+1}^{\ast ,j}\right\} _{j=1}^{B} = \left\{ 
	\widehat{r}_{h}^{\ast ,j}\left( \chi _{N+1}\right) \right\} _{j=1}^{B}.$$
	
	\item Draw $B$ i.i.d random variables $\left\{ \varepsilon
	_{N+1}^{\ast, j }\right\} _{j=1}^{B}$ from the empirical distribution
	of the centered residuals $\{\widehat{\epsilon}_{i,b}\}_{i \in S}$ (see (\ref{res})).
	
	\item For $j = 1, \dots, B$, obtain the future bootstrap observation 
	\begin{equation*}
		\zeta _{N+1}^{\ast, j} = \widehat{r}_{h}^{\ast ,j}\left( \chi _{N+1}\right) + \varepsilon _{N+1}^{\ast, j }.
	\end{equation*}
	
	\item Obtain the $C = [(1 - \alpha)B]$ deepest bootstrap future observations $\{ \zeta _{N+1}^{\ast, (j)}  \}_{j = 1}^C$ where $\zeta _{N+1}^{\ast, (j)}$ is the $j$-th deepest curve.
	
	\item Compute the lower and upper limits of the prediction region:
	$$\text{L}_{N+1}(t) = \min \{ \zeta _{N+1}^{\ast, (j)} (t)  : j = 1, \dots, C\}$$
	and
	$$\text{U}_{N+1}(t) = \max \{ \zeta _{N+1}^{\ast, (j)} (t): j = 1, \dots, C \}, $$
	respectively.
\end{enumerate}

\subsection{Prediction methods for functional time series}
\label{sec:Prediction_regions_forecasting}





Recalling the general formulation of the regression model given in (\ref{general_model}), different particular models can be assumed for the unknown regression function $r(\cdot)$ and different estimation methods 
can be considered. In this paper, two regression models with functional response will be used: the Functional Non Parametric (FNP) model and the Semi-Functional Partial Linear (SFPL) model. Both models have been used in \cite{Aneiros2013} for the functional prediction of residual demand curves in the electricity market and in \cite{Aneiros2016} for predicting daily electricity demand and price curves. For the case of the FNP model, Nadaraya-Watson type estimators were used; for the case of the SFPL model, a combination of least squares type estimators and Nadaraya-Watson type estimators were considered. Those models and estimators will be also used in this paper, and they are presented in the two following sections.

\subsubsection{The functional non parametric model}

Consider that the vector of explanatory variables in the general model (\ref{general_model}) is reduced to $\chi _{i}=\zeta _{i-1}$ (that is, no exogenous variables are considered). Then, model \eqref{general_model} becomes the functional autoregressive model of order 1
\begin{equation}
	\zeta _{i}=r(\zeta _{i-1})+\varepsilon _{i}, \quad i \in S.
	\label{FNP_model}
\end{equation}
Given that we only assume general conditions on $r(\cdot )$ (for instance, smoothness), (\ref{FNP_model}) is, in particular, a FNP model. Therefore, the operator $r(\cdot)$ should be estimated from some nonparametric approach. 
We will use the nonparametric Nadaraya-Watson type estimator, which is defined as follows:
\begin{equation}
	\widehat{r}_{h}^{FNP}\left( \zeta \right) =\sum\nolimits_{i\in
		S}w_{h}(\zeta ,\zeta _{i-1})\zeta _{i},
	\label{NW_estimator}
\end{equation}
where $w_{h}(\cdot,\cdot)$ are the Nadaraya-Watson weights given by
\begin{equation*}
	w_{h}(\zeta ,\zeta _{i-1})=\frac{K\left( d(\zeta,\zeta _{i-1})/h\right) 
	}{\sum\nolimits_{j\in S}K\left( d(\zeta ,\zeta _{j-1})/h\right) },
\end{equation*}
with $K:[0,\infty )\rightarrow \lbrack 0,\infty )$ being a kernel function and $h>0$ a smoothing parameter. In addition, $d(\cdot ,\cdot )$ denotes the semi-metric used to measure the proximity between two curves in $\mathcal{F}$. Then, a one-step-ahead prediction for the curve ${\zeta} _{N+1}$ can be obtained by means of $\widehat{\zeta} _{N+1}=\widehat{r}_{h}^{FNP}\left( \zeta _{N}\right)$.

Studies on the asymptotic properties of the estimator \eqref{NW_estimator} can be seen in \cite{Ferraty2012} and \cite{Zhu2017}, among others. In \cite{Aneiros2016} the model (\ref{FNP_model}) and the estimator \eqref{NW_estimator} are used to predict daily curves of electricity demand and price. The case of scalar response, that is, considering $\zeta _{i}(t)$ as response variable in (\ref{FNP_model}) instead of $\zeta _{i}$, was dealt in \cite{Masry2005} and \cite{Delsol2009} among others. Finally, the case of independence between the response and explanatory variables in general FNP models was considered in \cite{Ferraty2011}. 

\subsubsection{The semi-functional partial linear model}

Consider now that $\chi _{i}=(\zeta _{i-1}, \mathbf{x}_{i})$ and 
\begin{equation}
	r(\chi _{i})= 	\mathbf{x}_{i}^{\top}\boldsymbol{\beta} + m(\zeta _{i-1}) \label{r_SFPL}
\end{equation}
in the general model (\ref{general_model}). This means, in addition to an endogenous functional variable, exogenous escalar ones are included in the model. Then, we have that
\begin{equation}
	\zeta _{i}=\mathbf{x}_{i}^{\top}\boldsymbol{\beta} + m(\zeta _{i-1})+\varepsilon _{i}, \quad i \in S.
	\label{SFPL_model}
\end{equation}
As in the previous section, we assume that $m(\cdot )$ is an unknown smooth operator and $\boldsymbol{\beta}=(\beta _{1},\ldots ,\beta _{p})^{\top}$ is a vector of unknown functional parameters. 
Model (\ref{SFPL_model}) is, in particular, a SFPL model 
that extends  the previous FNP one (\ref{FNP_model}) by adding linear effects of some exogenous scalar variables in the regression function. Thus, the predictions obtained from model (\ref{SFPL_model}) could improve the predictions obtained from model (\ref{FNP_model}).

Model (\ref{SFPL_model}) was proposed in \cite{Aneiros2013} to predict curves of residual demand in electricity markets. In order to put in practice such prediction, they also proposed estimators for $\boldsymbol{\beta}$ and $m(\cdot)$ based on a combination of least squares and Nadaraya-Watson type estimates. Specifically, such estimators are given by
\begin{equation}
	\widehat{\boldsymbol{\beta}}_{h}=(\widetilde{\mathbf{X}}_{h}^{T}%
	\widetilde{\mathbf{X}}_{h})^{-1}\widetilde{\mathbf{X}}_{h}^{T}\widetilde{\boldsymbol{\zeta}}_{h}
	\label{beta_estimator}
\end{equation}
and 
\begin{equation}
	\widehat{m}_{h}^{SFPL}\left( \zeta \right) =\sum_{i\in S} w_{h}(\zeta
	,\zeta _{i-1})\left( \zeta _{i}-\mathbf{x}_{i}^{T}\widehat{\boldsymbol{\beta}}_{h}\right),
	\label{NW_estimator_SFPL}
\end{equation}
respectively, where
\begin{equation}
	\widetilde{\mathbf{X}}_{h}=(\mathbf{I}-\mathbf{W}_{h})\mathbf{X}
\end{equation}
and
\begin{equation}
	\widetilde{\boldsymbol{\zeta}}_{h}=(\mathbf{I}-\mathbf{W}_{h})\boldsymbol{\zeta}, 
\end{equation}
with $\mathbf{W}_{h}=(w_{h}(\zeta _{i},\zeta _{j}))_{i+1,j+1 \in S}$, 
$\mathbf{X}=(\mathbf{x}_{i})^{\top}_{i \in S}=(x_{ij})_{\substack{ i \in S  \\ 1\leq j\leq p }}$ and 
$\boldsymbol{\zeta}=(\zeta _{i})_{i \in S}.$

Now, from (\ref{r_SFPL}), (\ref{beta_estimator}) and (\ref{NW_estimator_SFPL}), we have that 
$$\widehat{r}^{SFPL}_{h}\left(\mathbf{x}, \zeta \right)=\mathbf{x}^{\top}\widehat{\boldsymbol{\beta}}_{h} + \widehat{m}_{h}^{SFPL}\left( \zeta \right)$$ 
is a natural estimator for $r(\mathbf{x},\zeta)$ defined in (\ref{r_SFPL}) in the SFPL model given in (\ref{SFPL_model}). 
Finally, 
the one-step-ahead forecast of the curve $\zeta _{N+1}$ 
is given by 
$$\widehat{\zeta }_{N+1}=\widehat{r}^{SFPL}_{h}\left(\mathbf{x}_{N+1}, \zeta_N \right).$$

It is worth being noted that the estimators (\ref{beta_estimator}) and (\ref{NW_estimator_SFPL}) were studied from a theoretical point of view in \cite{Aneiros2008} in the case of scalar response (that is, considering $\zeta _{i}(t)$ as response variable in (\ref{SFPL_model}) instead of $\zeta _{i}$).

\subsection{Tuning parameters}
\label{tuning}

As usual when one deals with nonparametric estimation, our proposals depend on several tuning parameters that must be chosen in some appropriate way. This section is devoted to indicate how such parameters are chosen in our application to real data (see Section \ref{sec:Application}).

In the bootstrap algorithms introduced in Section \ref{sec:Prediction_regions_building}, an auxiliary bandwidth, $b$, is needed, as usual when working with bootstrap procedures in nonparametric regression. The bandwidth $b$ is used to construct the residuals to be resampled and was theoretically proven to be larger than the bandwidth $h$ used to smooth the bootstrap sample (see \cite{Ferraty2010} and \cite{Rana2016} for further details). In this paper, the choice of the bandwidth $b$ followed the ideas of \cite{Vilar2018} for prediction intervals. See also \cite{Ferraty2011} and \cite{Aneiros2016} for more details.

The $L_p$ method depends on the chosen norm $\|\cdot\|_p$ and the depth-based method depends on the depth measurement.
Regarding the norm to be used, the usual norms $\|\cdot\|_1$ and $\|\cdot\|_2$ allow us to mathematically define the prediction region and to check whether or not a given curve belongs to this region. The disadvantage of these norms is that they do not allow a graphical representation of the region. Choosing the norm $\|\cdot\|_{\infty}$ 
could provide a useful graphical representation of the resulting prediction regions. For this reason, in Section \ref{sec:Application} the norm $\|\cdot\|_{\infty}$ will be used. The method for obtaining prediction regions based on $L_{\infty}$ has the drawback of not taking into account the volatility of the functional curves. But this method will be compared with the other two proposals that do consider this volatility.
Concerning the choice of depth measurement, in Section \ref{sec:Application}, the depth-based method uses the Random Tukey depth proposed by \cite{Cuesta2008}, which is not very time consuming, since it is possible to obtain similar results with it to those obtained with more involved depths by taking only a few one-dimensional projections.

Finally, the forecasting procedures exposed in Section \ref{sec:Prediction_regions_forecasting} depend on various tuning parameters that must be selected from the data. Specifically, the FNP and SFPL predictors depend on the bandwidth $h$, the semi-metric $d(\cdot,\cdot )$ and the kernel function $K(\cdot )$. To choose $h$, $d(\cdot,\cdot )$ and $K(\cdot )$, the suggestions given in \cite{Aneiros2013} were followed. 
For the bandwidth, $h$, the $k$-nearest-neighbours method was considered. The number of neighbours, $k$, was selected by means of the local cross-validation method.
The semimetric, $d(\cdot ,\cdot )$, is chosen based on a seminorm. Since our data are rough, we consider a seminorm based on functional principal component analysis. 
Finally, we use the Epanechnikov kernel defined by $K(u)=0.75(1-u^{2})I_{(0,1)}$, although it is known that the choice of kernel has little influence on the results.

\section{The data}
\label{sec:Data}

Our goal is to compute prediction regions of the daily curves of electricity demand and price, one day-ahead, corresponding to mainland Spain, year 2012. Some information will be taken from past curves (endogeneus functional variables). In addition, exogeneus scalar covariates could be used. Specifically, our database contains information related to years 2011 and 2012: hourly observations of electricity demand and price, maximum daily temperature and daily wind power production.

This data set was previously analysed in various articles from different models and with different goals: \cite{Aneiros2016} obtained pointwise predictions for hourly electricity demand and price from FNP (\ref{FNP_model}) and SFPL (\ref{SFPL_model}) models, by changing the functional response $\zeta _{i}$ by the scalar one $\zeta _{i}(j), \ j=1,\ldots,24$; based on the same models, \cite{Vilar2018} completed such study by computing prediction intervals for future hourly electricity demand and price; \cite{Ranha2018} also reported both pointwise prediction and prediction intervals for hourly electricity demand and price, but based on additive models instead of FNP and SFPL ones; \cite{Aneiros2016}, in addition to the aforementioned pointwise (scalar) prediction, dealt the case of prediction of daily curves of electricity demand and prices from models (\ref{FNP_model}) and (\ref{SFPL_model}) (now, maintaning the functional response $\zeta _{i}$). To our knowledge, there are no analyses of this data set that present prediction regions for the future curves of daily electricity demand and price.

The motivation for choosing these data to illustrate the proposals of this work is twice: on the one hand, to be able to compare them with those proposed in \cite{Vilar2018} for prediction intervals; on the other hand, to complete and finalise the serie of papers related to the analysis of these data presented in the previous paragraph.


Each daily functional data comes from the 24 hourly observations of demand (measured in MWh, Megawatt-hour) or price (measured in Cent/kWh), in each day. Note that smoothing techniques are used to convert the 24 hourly data in a functional observation (a curve). Scalar covariates are related to daily demand, temperature and wind power production.

\begin{figure}[H]
	\centering
	\includegraphics[width=0.45\linewidth]{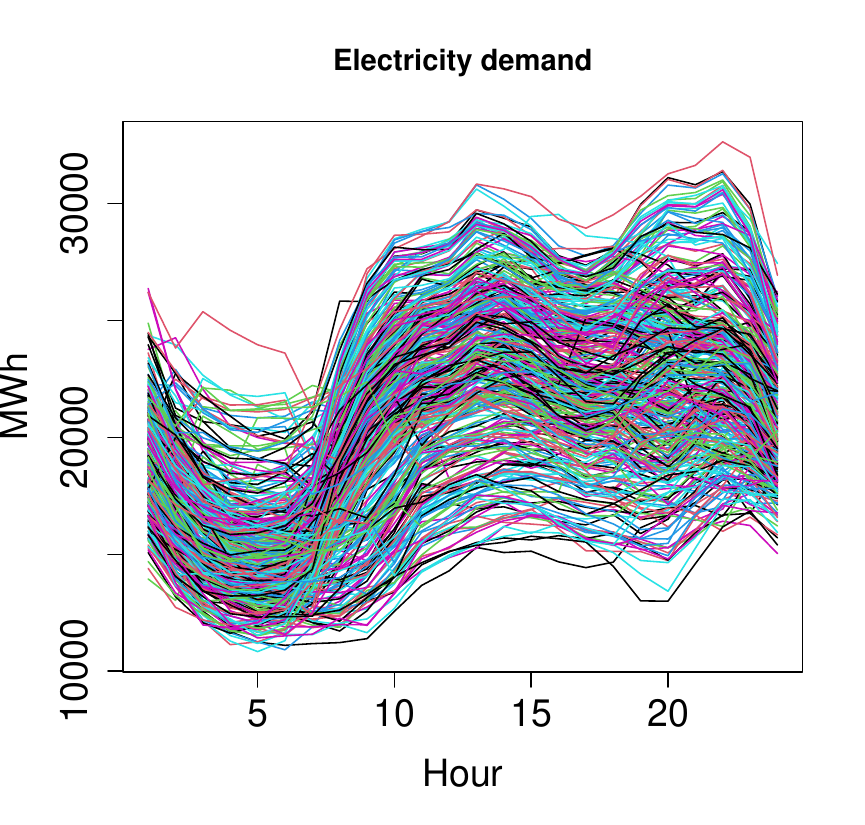}
	\includegraphics[width=0.45\linewidth]{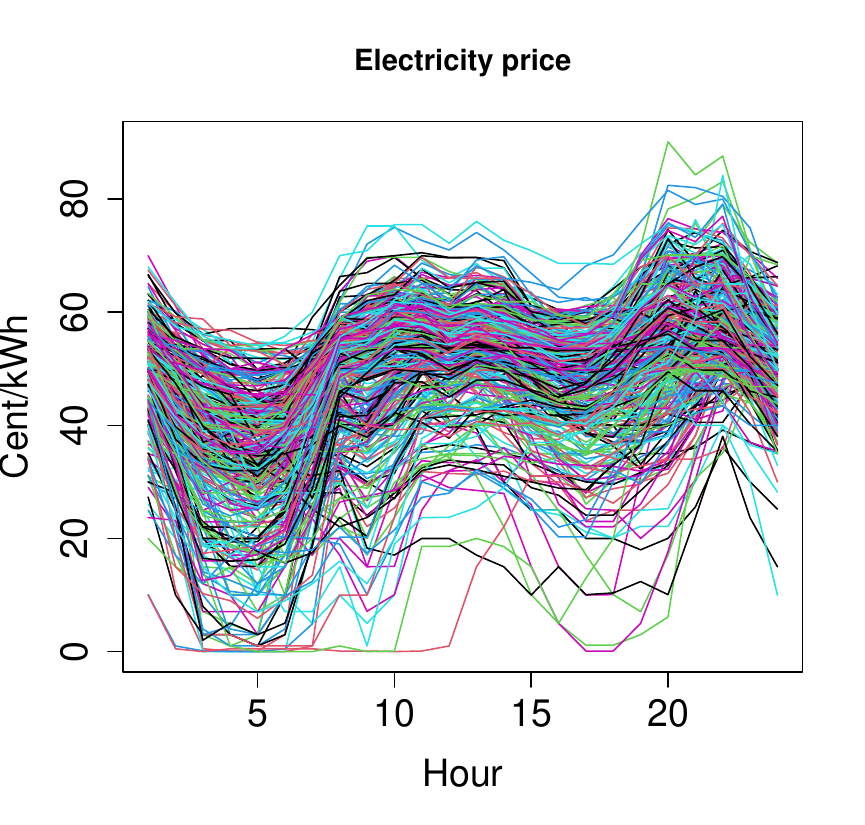}
	\caption{Electricity demand (left) and price (right) daily curves along the year 2012.}
	\label{fig:Demand_price}
\end{figure}

\begin{figure}[H]
	\centering
	\includegraphics[width=1\linewidth]{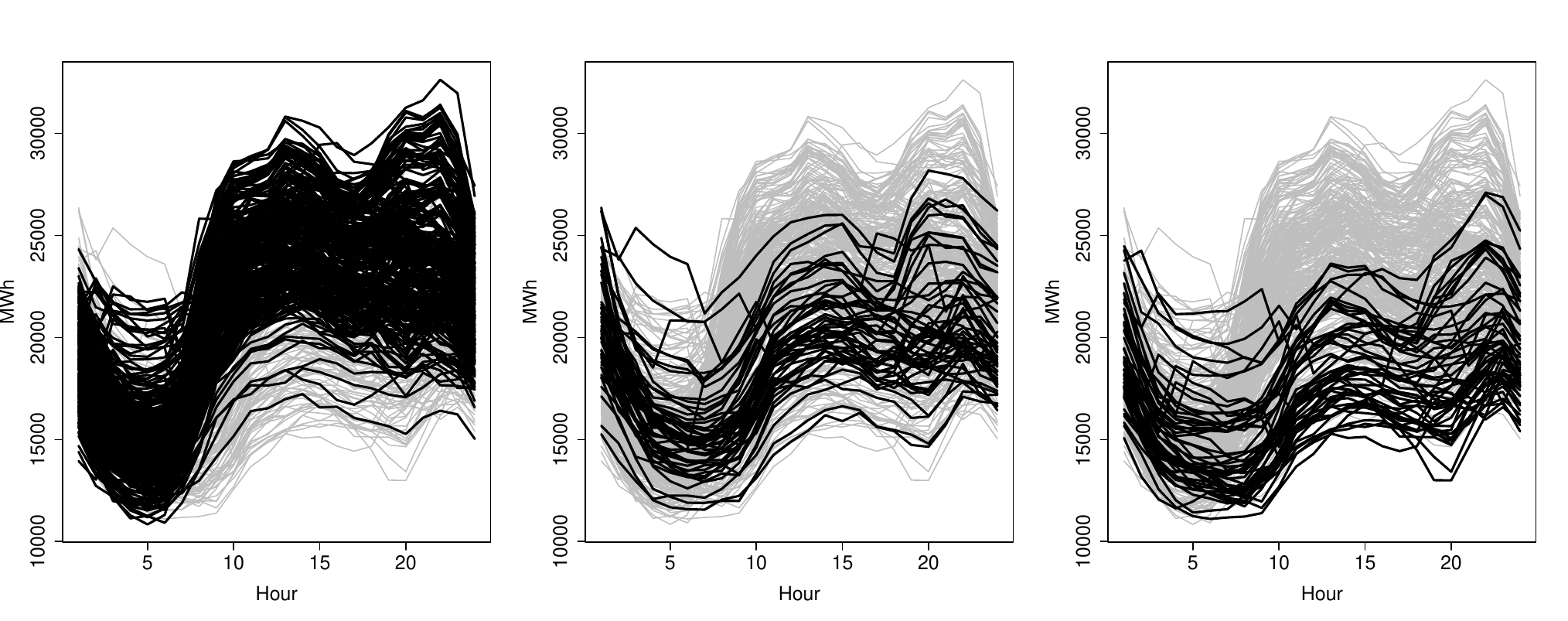}
	\caption{Electricity demand daily curves along the year 2012 distinguishing weekdays (left), Saturdays (middle) and Sundays (right).}
	\label{fig:Demand_days}
\end{figure}

\begin{figure}[H]
	\centering
	\includegraphics[width=1\linewidth]{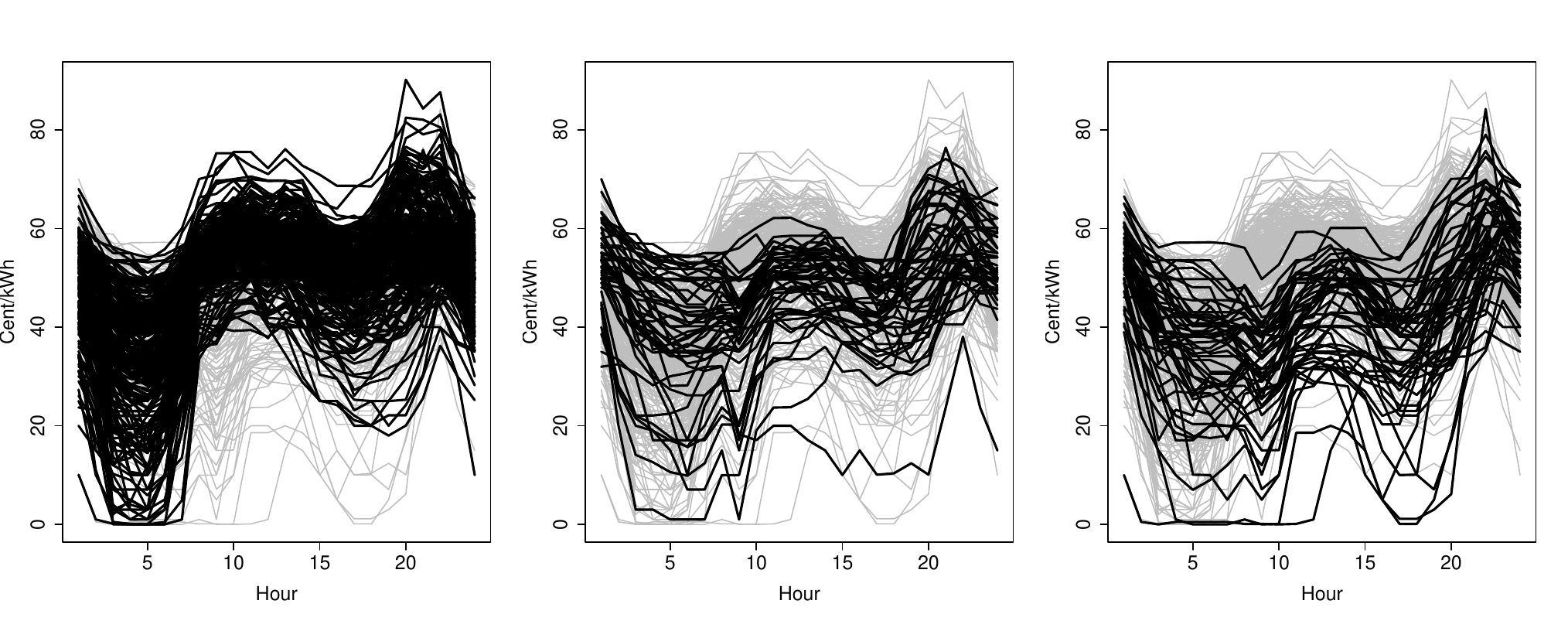}
	\caption{Electricity price daily curves along the year 2012 distinguishing weekdays (left), Saturdays (middle) and Sundays (right).}
	\label{fig:Price_days}
\end{figure}

Electricity demand and price curves show daily and weekly seasonality and are affected by the weekend calendar. Figure \ref{fig:Demand_price} shows the daily electricity demand and price curves along the year 2012.
Figures \ref{fig:Demand_days} and \ref{fig:Price_days} show the differences in electricity demand and price, respectively, according to the day type: weekday (Monday to Friday), Saturday and Sunday. 
A more comprehensive analysis of this effect can be found in \cite{Vilar2018}. Due to this feature detected in the curves, the analyses carried out in this work will be done separately for each type of day.

Both demand and price daily curves present some outliers. To identify these atypical observations, methods proposed in \cite{Aneiros2015} and \cite{Vilar2016} are used. In \cite{Aneiros2015} a test based on functional depths is performed to determine if a curve is an outlier or not. In \cite{Vilar2016}, the authors consider robust principal component analysis to look for outliers in the projection of the first component. The identified outliers are replaced in the sample by a weighted moving average of the surrounded days. See \cite{Vilar2018} for more details.

There exist exogeneus variables that could improve the forecasts and, therefore, the prediction regions of the daily curves of electricity demand and price. See \cite{Taylor2003}, \cite{Taylor2006} and  \cite{Hyde2015}, among others, for a discussion about the impact of meteorological factors on the electricity demand and the effect of the electricity demand and wind power production on the electricity price.
In \cite{Aneiros2016} and \cite{Vilar2018}, a detailed analysis of these data and the influence of certain covariates on them is carried out.

In general, there is a high demand for electric heating in cold weather. Thus, the temperature influences the electricity demand and the maximum daily temperature (in Celsius degrees, $^{\circ}C$) in Spain is considered for the study. However, this variable has a nonlinear effect on demand; demand as a function of temperature exhibits a U-shaped pattern.
Since the prediction methods to be used in this study (see Section \ref{sec:Prediction_regions_forecasting}) include the information of the exogenous variables in a linear way, some transformation of the maximum temperature that verifies this assumption is needed. 
In this paper, we consider the HDD (Heating Degree Days) and CDD (Cooling Degree Days) variables, which exert a linear effect on the electricity demand. They are defined as follows:
$$ \text{HDD}(t) = \max\{ 20 - T(t), 0 \} $$
and
$$ \text{CDD}(t) = \max\{ T(t)  - 24, 0 \} .$$
where $T(t)$ is the maximum daily temperature in day $t$. See \cite{Cancelo2008} and \cite{Aneiros2016} for more details on the justification behind the construction of these variables.

For the building of prediction regions of the daily electricity price curves, both daily forecasted demand and wind power production will be considered as scalar covariates. The System Operator in the Spanish Electricity Market, REE, monitors the demand and its generator structure and provides the amount of demand covered by wind power during each ten minutes period of the year. Then, the hourly or daily demand covered by wind power can be obtained. In this case, both covariates do have a linear effect on the electricity price.

It is important to note that unobserved covariates (temperature and wind production for the day to be forecast) are going to be included in the regression models. On the one hand, in order to apply the SFPL procedure, it is necessary to have accurate forecasts of the values of these covariates for the following day. On the other hand, sophisticated meteorological models are known to provide very good forecasts of temperature and wind power. However, neither these models nor these forecasts are publicly available. Since the predictive power of the SFPL model could be greatly diminished or masked if not very good forecasts of these covariates are incorporated in the model, we chose to include the ideal forecasts given by the values themselves.

\section{Application to the electricity market}
\label{sec:Application}

In this section, prediction regions for the daily curves of electricity demand and price are computed using the three bootstrap methods presented in Section \ref{sec:Prediction_regions_building}: $L_p$-method, $\lambda$-method and depth-based method. Although the methods can be extended to any forecasting method, in this section we will use those associated with the FNP and SFPL models presented in Section \ref{sec:Prediction_regions_forecasting} (see (\ref{FNP_model}) and (\ref{SFPL_model}), respectively).

Electricity demand and price are both considered to be continuous time
stochastic processes with seasonal length $\tau=24$, and the notation $\left\{ \zeta \left(
t\right) \right\} _{t\in R}$ is used to refer any of them (units for $t$ are hours). 
The aim of this section is to put in practice the prediction regions proposed in Sections \ref{Lp-method}, \ref{lambda-method} and \ref{depth-method}. More specifically, to compute prediction regions for each daily curve, $\zeta_{N+1}$, of electricity demand or price in year 2012 from information given by the previous 365 days. This set of curves corresponding to the 365 curves preceding $\zeta_{N+1}$ is determined by the following set of index: $\mathcal{I} = \{ N - 364, N - 363, \dots, N-1, N \}$.
However, the dynamic of the curves of demand or price depends on the type of day to be predicted: weekdays, Saturdays or Sundays (see Figures \ref{fig:Demand_days} and \ref{fig:Price_days}).
This fact suggests to compute three different regression models according to the type of day: to forecast a curve corresponding to either Sunday or Saturday, information from the previous curve (i.e. from the curve observed on previous Saturday or Friday, respectively) will be used; meanwhile, if one wishes to forecast a curve corresponding to a weekday, information will be taken from the curve observed on the previous weekday (note that the previous weekday to a Monday is a Friday). 
Therefore, the following notation is introduced to distinguish the various scenarios:
\begin{itemize}
	\item $S_{\text{W}} =\{1,\ldots,\#(S^0_{\text{W}}) \},$ where $S^0_{\text{W}} = \{ i \in \mathcal{I} : \;  i \text{ is a weekday}\}$ is the set of indices corresponding to weekdays in the 365 curves preceding $\zeta_{N+1}$. In addition we denote, $\overline{\zeta}_{i}=\zeta _{(i)}$, for $i \in S_{\text{W}}$, where $(i)$ is the element in $S^0_{\text{W}}$ whose rank is $i$, 
	
	\item $S_{\text{Sat}} = \{ i \in \mathcal{I} : \;  i \text{ is a Saturday}\}$ is the set of indices corresponding to Saturdays in the 365 curves preceding $\zeta_{N+1}$,
	\item $S_{\text{Sun}} = \{ i \in \mathcal{I} : \;  i \text{ is a Sunday}\}$ is the set of indices corresponding to Sundays in the 365 curves preceding $\zeta_{N+1}$.
\end{itemize}
Note that the notation above aims to be able to use, in each of the three scenarios, our notation corresponding to the algorithms for prediction regions proposed in Section \ref{sec:Prediction_regions}. In the case that $\zeta_{N+1}$ corresponds to Saturday or Sunday, it is suffice to consider $S=S_{\text{Sat}}$ or $S=S_{\text{Sun}}$, respectively. If $\zeta_{N+1}$ corresponds to a weekday, $S=S_{\text{W}}$ should be considered and, in addition, curves $\{\overline{\zeta}_{i}\}$ should be used instead of $\{{\zeta}_{i}\}$. Note also that the number of curves available to predict $\zeta_{N+1}$ depends on the type of day, since the sets $S_{\text{W}}$, $S_{\text{Sat}}$ and $S_{\text{Sun}} $ do not have the same cardinal. Specifically, $\#(S_{\text{W}}) = 261$, $\#(S_{\text{Sat}}) = 52$ and $\#(S_{\text{Sun}})= 53$. Therefore, the sample size to estimate the daily demand or price curve is much larger if one wants to estimate a day of the weekday type than if it is Saturday or Sunday.

In this study, as mentioned in Section \ref{tuning}, $L_p$ method uses the norm $L_{\infty}$ and the depth-based method uses the random Tukey depth. The random Tukey depth is available at the \textsl{fda.usc} package from the Comprehensive R Archive Network (see \cite{fda.usc}). The number of resamples for the bootstrap procedures is $B=500$.

\subsection{Performance measurements}

A prediction region performs well if its coverage is close to the nominal one, $\left( 1-\alpha \right) 100\%$, and has a small area or average width. The following values measure the performance of a set of $J$ prediction regions (one prediction region for each curve $\zeta_{N_j+1}$, $j=1,\ldots,J$) and allow for the comparison of results.

Denoting $\text{L}_{N_j+1}$ the lower limit and $\text{U}_{N_j+1}$ the upper limit of the prediction region for $\zeta_{N_j+1}$, with $j=1,\ldots,J$, the functional coverage is the percentage of bootstrap regions that contain to the corresponding daily curves of electricity demand or price, and it is defined as follows:
\begin{equation*}
	\label{eq:Coverage}
	FCov= 100 \times \dfrac{1}{J} \sum_{j = 1}^{J} I\big( \zeta_{N_j+1}(t) \in \big(\text{L}_{N_j+1}(t), \text{U}_{N_j+1}(t) \big), \, \forall t \in (0, 24]\big).
\end{equation*}

The mean pointwise coverage is the mean of the percentage of time grid values for which the prediction regions contain the corresponding daily curves of electricity demand or price. It is given by
\begin{equation*}
	\label{eq:Pointwise_coverage}
	PCov= 100 \times \dfrac{1}{J} \sum_{j = 1}^{J}  \bigg(   \dfrac{1}{24}  \sum_{t=1}^{24}  I\big( \zeta_{N_j+1}(t) \in \big(\text{L}_{N_j+1}(t), \text{U}_{N_j+1}(t) \big)   \bigg).
\end{equation*}

Average width of the bootstrap prediction region is defined by
\begin{equation*}
	\label{eq:Width}
	AWidth= \dfrac{1}{J} \sum_{j = 1}^{J} \Bigg(   \dfrac{1}{24}  \sum_{t=1}^{24} \big( \text{U}_{N_j+1}(t) - \text{L}_{N_j+1}(t)   \big) \Bigg).
\end{equation*}

Winkler or interval score (see \cite{Winkler1972} and \cite{Geneiting2007}) is also used to compare the behaviour of the methods. For classical prediction intervals, it is defined as the length of the interval plus a penalty if the theoretical value is outside the interval. Thus, it combines width and coverage. For values that fall within the interval, the Winkler score is simply the length of the interval. So, low scores are associated with narrow intervals. When the theoretical value falls outside the interval, the penalty is proportional to how far the observation is from the interval. Given a prediction interval, $(\text{L}_{N+1}(t),\text{U}_{N+1}(t))$, for $\zeta_{N+1}(t)$ ($t$ is fixed), the formula of the Winkler score is as follows:
\begin{equation*}
	\begin{array}{rcl}
		\text{WS}_{N+1} (t)  & = & \text{U}_{N+1}(t) -\text{L}_{N+1}(t) 
		+ \frac{2}{\alpha} (\text{L}_{N+1}(t) - \zeta_{N+1}(t)) 
		I\big(  \zeta_{N+1}(t) < \text{L}_{N+1}(t) \big) 
		\\
		& & 
		+ \frac{2}{\alpha} (\zeta_{N+1}(t) - \text{U}_{N+1}(t)) 
		I\big(  \zeta_{N+1}(t) > \text{U}_{N+1}(t) \big).
	\end{array}
	\label{eq:WS}
\end{equation*}
Since we are working with prediction regions, the functional Winkler score is proposed as a criteria for the comparison. It is an extension of the classic Winkler score to the functional context, defined as follows:
\begin{equation*}
	\begin{array}{rcl}
		\text{FWS}_{N+1} & = & \delta(\text{L}_{N+1} , \text{U}_{N+1})
		\\
		&  & + \frac{2}{\alpha} \min \big\{  \delta (    \text{L}_{N+1}, \zeta_{N+1}  ) ,  \delta (    \text{U}_{N+1}, \zeta_{N+1}  ) \big\}
		\\
		& & I\big( \exists t \in (0, \tau] : \zeta_{N+1}(t)  < \text{L}_{N+1}(t) \text{ or } \zeta_{N+1}(t)  > \text{U}_{N+1}(t) \big),
	\end{array}
	\label{eq:FWS}
\end{equation*}
where $\delta(\cdot,\cdot )$ is a semi-metric. Then, when one has several prediction regions, one considers the mean of their functional Winkler scores: 
$$\text{FWS}=\dfrac{1}{J} \sum_{j=1}^J \text{FWS}_{N_j+1}.$$
In our study, the semi-metric to be considered is 
\begin{equation}
	\delta(\xi,\chi )=\left\Vert \xi-\chi \right\Vert _{1} = \int_{(0,\tau]}\left\vert \xi (t) - \chi (t) \right\vert dt.
\end{equation}

Section \ref{sec:Application_demand} contains the results for the electricity demand curves. Section \ref{sec:Application_price} contains an analogous study for the electricity price curves.

\subsection{Prediction regions for the electricity demand daily curves}
\label{sec:Application_demand}

Previous works have proposed methods to obtain prediction intervals for hourly electricity demand and price. In \cite{Vilar2018}, prediction intervals are obtained for each hour, one-day ahead, of the electricity demand and price, $ \zeta _{N+1} (t)$ with $t \in \{  1, \dots, 24 \}$, in the year 2012. They proved the good performance of the prediction intervals in the problem of the next-day forecasting of electricity demand and price. Table \ref{tab:PR_demand_FNP_comparison_FPI} contains a comparison between the methods proposed in this paper and the use of the prediction intervals from \cite{Vilar2018} as prediction regions (after smoothing the former) for the electricity demand curves in the year 2012. The functional non parametric forecasting method is considered for this analysis.

The results in Table \ref{tab:PR_demand_FNP_comparison_FPI} show that the functional coverage of the prediction regions obtained by smoothing prediction intervals (PI) is far from the nominal coverage: for a nominal coverage of $95\%$, the method based on PI provides functional coverages of around $70\%$, down even to $54\%$. 
It is also observed that the functional Winkler score of the method based on pointwise prediction intervals is higher than in the three methods for functional prediction regions. 
This is to be expected since computing a pointwise prediction interval for each of the 24 points considered leads to narrow intervals. But globally, they constitute a narrow region with a high probability that some point of the real curve is not covered. 
Therefore, their use as a method for computing functional prediction regions is discarded.

\begin{table}[H]
	\centering
	\begin{tabular}{cc|c|c|c|c|} 
		\cline{3-6}
		\multicolumn{1}{l}{}                                                           & \multicolumn{1}{l|}{} & Weekdays     & Saturdays & Sundays & Year \\ \hline
		\multicolumn{1}{|c|}{\multirow{1}{*}{$L_{\infty}$-method}} 
		& FCov
		& 93.8 & 90.4  &  84.9 & 92.0 \\ \cline{2-6}
		\multicolumn{1}{|c|}{}                                                         & PCov    
		&  98.8 &    97.5  & 97.4 & 98.4   \\ \cline{2-6}
		\multicolumn{1}{|c|}{}                                           & AWidth   
		& 8360.0  & 11594.6  &  10294.8 & 9099.7  \\ \cline{2-6}
		\multicolumn{1}{|c|}{}                                         & FWS  
		& 13433.3 & 15399.1 & 16896.6  & 14214.1 \\ 
		\hline
		\multicolumn{1}{|c|}{\multirow{1}{*}{$\lambda$-method}}  
		& FCov   
		& 95.4 &  90.4   &  81.1  &  92.6  \\ \cline{2-6}
		\multicolumn{1}{|c|}{}                                                         & PCov    
		& 99.2 &  98.2   & 96.5 &   98.7 \\ \cline{2-6}
		\multicolumn{1}{|c|}{}                                           & AWidth   
		&  7684.8 & 11272.5  &  9414.3 & 8444.9  \\ \cline{2-6}
		\multicolumn{1}{|c|}{}                                         & FWS  
		& 10684.2 & 14957.7 & 16970.9 & 12201.7 \\ 
		\hline
		\multicolumn{1}{|c|}{\multirow{1}{*}{Depth method}}  
		& FCov
		& 92.3 & 82.7  & 69.8   &  87.7  \\ \cline{2-6}
		\multicolumn{1}{|c|}{}                                                         & PCov    
		& 98.6 &  97.0 &  93.4  & 97.6    \\ \cline{2-6}
		\multicolumn{1}{|c|}{}                                           & AWidth   
		& 6932.8  &  9128.5 & 7780.8  &  7367.5    \\ \cline{2-6}
		\multicolumn{1}{|c|}{}                                         & FWS  
		& 13014.0 & 20035.7  &   27992.8 &  16180.7  \\ 
		\hline
		\multicolumn{1}{|c|}{\multirow{1}{*}{PI}}  
		& FCov
		& 70.1  & 53.8 &  64.2 & 66.9     \\  \cline{2-6}
		\multicolumn{1}{|c|}{}  
		& PCov    
		& 93.7 & 88.2 &  90.4 &  92.4  \\  \cline{2-6}
		\multicolumn{1}{|c|}{}                                           & AWidth  
		& 5339.4 & 4384.5 & 5389.3 & 5211.0  \\ \cline{2-6}
		\multicolumn{1}{|c|}{}                                           & FWS  
		& 21005.0  &  24717.9 & 25680.9 & 22209.7  \\ 
		\hline
	\end{tabular}
	\caption{Functional coverage (in $\%$), pointwise coverage, width and functional Winkler score of the prediction regions for the daily electricity demand based on the FNP model using the $L_{\infty}$-method, the $\lambda$-method, the depth-based method and the prediction regions obtained by smoothing the prediction intervals (PI) with $\alpha = 0.05$ for each type of day in 2012.}
	\label{tab:PR_demand_FNP_comparison_FPI}
\end{table}

Table \ref{tab:PR_demand_FNP_week_sat_sun_80} shows the coverage, width and Winkler score of the prediction regions obtained with the proposed methods and using the FNP forecasting model for the 2012 electricity demand curves at $80\%$ confidence level. The condition that the entire daily curve is contained in the prediction region is a demanding one, so working with high confidence levels leads to very wide and, in many cases, uninformative prediction regions. Therefore, it is reasonable to work with $\alpha=0.20$ in this context. Results for $95\%$ were already shown in Table \ref{tab:PR_demand_FNP_comparison_FPI}.

The coverage percentages of the three methods are reasonable for weekdays, but they are low on Saturdays and Sundays, especially on the latter. At the same time, the forecast regions obtained for Saturdays and Sundays are wider than for weekdays.
This performance is motivated by the sample size of $\mathcal{I}_W$, $\mathcal{I}_{Sat}$ and $\mathcal{I}_{Sun}$, that is, the number of preceding curves to be used in the prediction for each group. 
When predicting demand curves on Saturdays and Sundays, the sample size of $\mathcal{I}_{Sat}$ and $\mathcal{I}_{Sun}$, respectively, is much smaller than the sample size of $\mathcal{I}_W$ for weekdays.
In addition, Sunday's daily electricity demand curves are highly variable and make the problem of forecasting and construction of prediction regions difficult.

In the comparison of the three methods it is observed that $\lambda$-method provides the smallest FWS. Furthermore, this method has the advantage of taking into account the volatility of the demand curve.
The other two methods present similar results, slightly better for the $L_{\infty}$-method whose drawback is providing confidence regions of constant width throughout the day.

\begin{table}[H] 
	\centering
	\begin{tabular}{cc|c|c|c|c|}
		\cline{3-6}
		\multicolumn{1}{l}{}                                             & \multicolumn{1}{l|}{} & Weekdays & Saturdays   & Sundays & Year \\ \hline
		\multicolumn{1}{|c|}{\multirow{1}{*}{$L_{\infty}$-method}} 
		& FCov
		& 77.8 &  80.8  &  66.0  & 76.5  \\ \cline{2-6}
		\multicolumn{1}{|c|}{}                                           & AWidth   
		& 5640.9 &  7570.4 & 7231.6   & 6145.4  \\ \cline{2-6}
		\multicolumn{1}{|c|}{}                                         & FWS  
		& 8937.3  & 10006.1 & 12278.0 & 9572.9 \\ 
		\hline
		\multicolumn{1}{|c|}{\multirow{1}{*}{$\lambda$-method}}  
		& FCov
		& 73.9 &  76.9  &  66.0  & 73.2 \\ \cline{2-6}
		\multicolumn{1}{|c|}{}                                           & AWidth   
		&  5285.8 &   7084.4 &  6708.9  & 5747.4  \\ \cline{2-6}
		\multicolumn{1}{|c|}{}                                         & FWS  
		& 8798.3 & 10234.9   &  11586.0  & 9406.1 \\ 
		\hline
		\multicolumn{1}{|c|}{\multirow{1}{*}{Depth method}}  
		& FCov
		& 76.2 &  76.9 & 52.8   &  72.9 \\ \cline{2-6}
		\multicolumn{1}{|c|}{}                                           & AWidth   
		& 5851.9 & 7955.4   & 5750.0   &  6136.0   \\ \cline{2-6}
		\multicolumn{1}{|c|}{}                                         & FWS  
		& 9212.4  & 11140.7 & 12276.3 & 9930.1 \\ 
		\hline
	\end{tabular}
	\caption{Functional coverage (in $\%$), width and functional Winkler score of the prediction regions for the daily electricity demand based on the FNP model using the $L_{\infty}$-method, the $\lambda$-method and the depth-based method at $80\%$ confidence level for each type of day in 2012.}
	\label{tab:PR_demand_FNP_week_sat_sun_80}
\end{table}

\begin{figure}[H]
	\centering
	\includegraphics[width=0.4\linewidth]{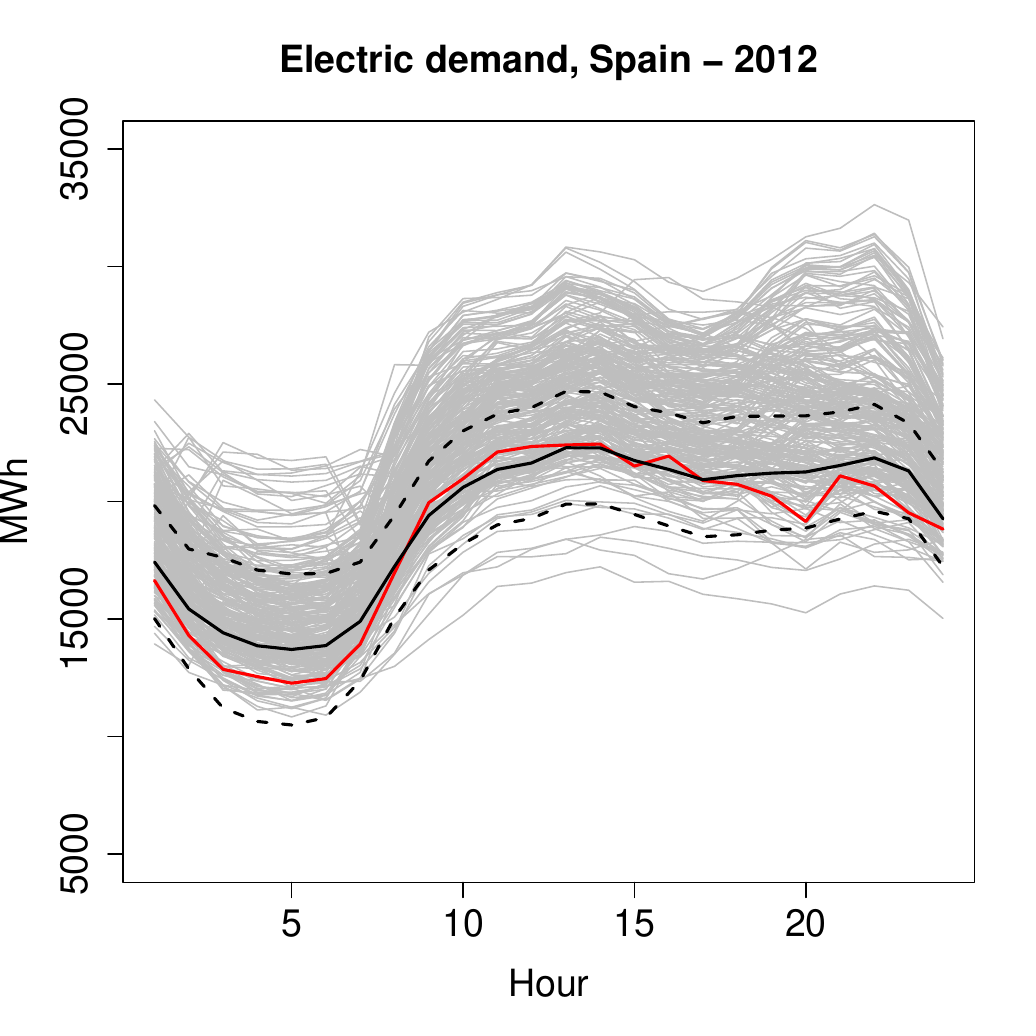}
	\includegraphics[width=0.4\linewidth]{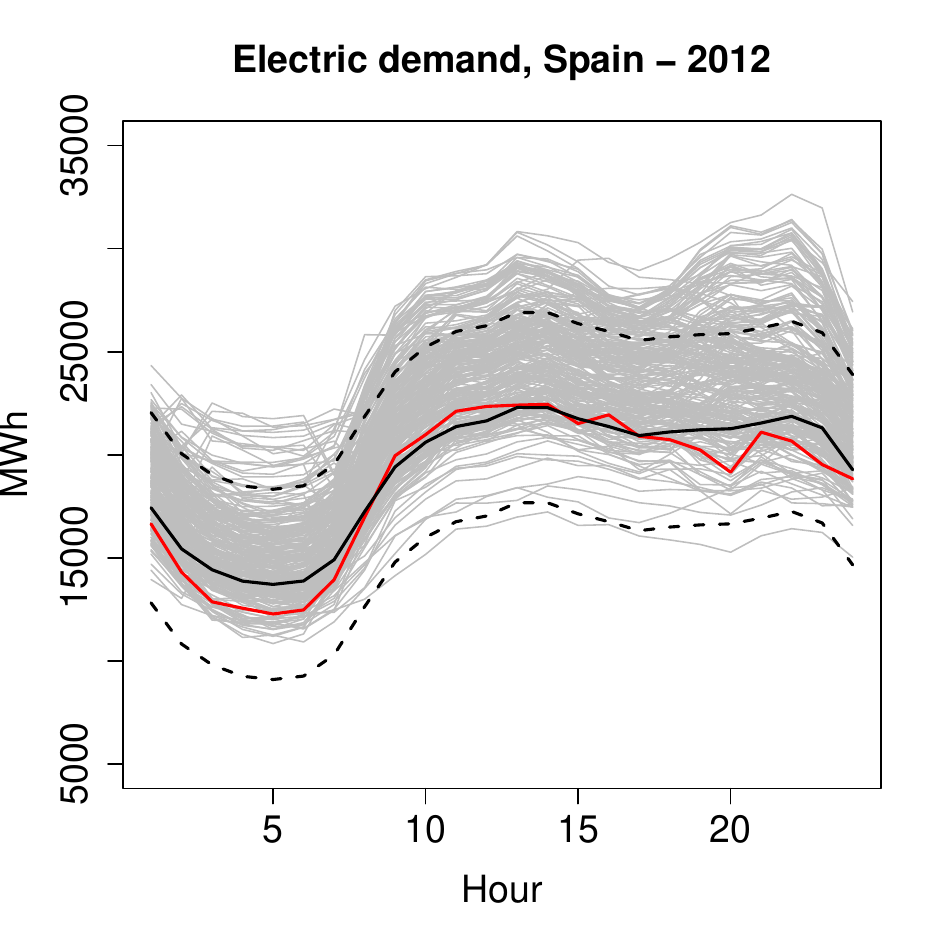}
	\caption{Daily curves of electricity demand corresponding to weekdays (grey solid lines), curve of electricity demand for 2nd April 2012 (red solid line), prediction curve (black solid line) and prediction region (black dashed lines) by means of $\lambda$-method at $80\%$ of confidence (left panel) and $L_{\infty}$-method at $95\%$ of confidence  (right panel) based on the FNP forecasting model.}
	\label{fig:PR_demand_FNP_weekdays}
\end{figure}

Figure \ref{fig:PR_demand_FNP_weekdays} shows the forecast and prediction region of electricity demand on a weekday using the $\lambda$-method at the $80\%$ confidence level and the $L_{\infty}$-method at the $95\%$ confidence level based on the FNP model. The prediction region obtained at $80\%$ is more informative, as it has a smaller width. The $\lambda$-method also takes into account the volatility of electricity demand throughout the day. 
On the contrary, the $L_{\infty}$-method provides a region of larger width because the $95\%$ confidence level and also constant for each hour.

The methods for obtaining prediction regions for the electricity demand daily curves based on the SFPL regression model are discussed below. 
In this work, the vector of scalar covariates included in the SFPL model given in \eqref{SFPL_model} to forecast the daily curves of electricity demand is $\mathbf{x}=(x_1,x_2)^T=(\text{HDD},\text{CDD})^T$, where HDD and CDD are the Heating Degree Days and Cooling Degree Days, respectively, introduced in Section \ref{sec:Data}.

Table \ref{tab:PR_demand_SFPL_week_sat_sun} shows the functional coverage, width and Winkler score of the prediction regions obtained with the proposed methods, and using the SFPL forecasting model, for the 2012 electricity demand curves at confidence levels $80\%$ and $95\%$. 

The results obtained with the SFPL regression model are very similar to those obtained with the FNP model. Comparing Tables \ref{tab:PR_demand_FNP_comparison_FPI}, \ref{tab:PR_demand_FNP_week_sat_sun_80} and \ref{tab:PR_demand_SFPL_week_sat_sun}, it is concluded that the functional coverage of the $80\%$ prediction regions based on the SFPL regression model is lower than that of the regions based on the FNP, but so is the region width, which leads to lower FWS. When working with a nominal coverage of $95\%$ the effect is the opposite.

Regarding the methods for obtaining prediction regions, the worst behaviour with lower coverages and larger widths is observed for the depth-based method. The $L_{\infty}$ and $\lambda$ methods provide similar results, slightly better for the $L_{\infty}$-method.

\begin{table}[H]
	\centering
	\begin{tabular}{cc|c|c|c|c|}
		\hline
		\multicolumn{2}{|c|}{$\alpha = 0.05$}  & Weekdays & Saturdays   & Sundays & Year \\ \hline
		\multicolumn{1}{|c|}{\multirow{1}{*}{$L_{\infty}$-method}} 
		& FCov    
		& 90.8 & 88.5  &  79.2 & 89.0 \\ \cline{2-6}
		\multicolumn{1}{|c|}{}                                           & AWidth   
		& 8469.0 & 10318.0  &  8700.9 &  8765.3  \\ \cline{2-6}
		\multicolumn{1}{|c|}{}                                           & FWS   	
		& 16084.8 & 19116.7  & 23472.6 & 17585.4 \\ 
		\hline
		\multicolumn{1}{|c|}{\multirow{1}{*}{$\lambda$-method}}  
		& FCov 
		& 90.8 &  86.5 & 69.8 & 87.1 \\ \cline{2-6}
		\multicolumn{1}{|c|}{}                                           & AWidth   
		&  7793.2 &  9845.6 & 7609.6 & 8058.2 \\ \cline{2-6}
		\multicolumn{1}{|c|}{}                                           & FWS   		
		& 14724.5 & 19648.2  & 27294.4 & 17244.3  \\
		\hline
		\multicolumn{1}{|c|}{\multirow{1}{*}{Depth method}}  
		& ACoverage
		& 87.0 & 76.9 & 62.2 &  82.0 \\ \cline{2-6}
		\multicolumn{1}{|c|}{}                                           & AWidth   
		& 7526.2 &  8039.3 & 6967.7 & 7518.2 \\ \cline{2-6}
		\multicolumn{1}{|c|}{}                                           & FWS   		
		& 16398.0 & 20954.7  &  29681.6 & 18969.0  \\   \hline
		\multicolumn{2}{|c|}{$\alpha = 0.20$}  & Weekdays & Saturdays   & Sundays & Year\\ \hline
		\multicolumn{1}{|c|}{\multirow{1}{*}{$L_{\infty}$-method}} 
		& FCov
		& 74.3  & 80.8 & 56.6 & 72.7 \\ \cline{2-6}
		\multicolumn{1}{|c|}{}                                           & AWidth   
		& 5284.8 & 6824.9 & 6171.3 & 5632.0 \\ \cline{2-6}
		\multicolumn{1}{|c|}{}                                           & FWS   	
		& 8800.6 &  9213.6 & 11820.8 &  9296.6  \\ 
		\hline
		\multicolumn{1}{|c|}{\multirow{1}{*}{$\lambda$-method}}  
		& FCov
		& 71.3 & 75.0 & 47.2 & 68.3 \\ \cline{2-6}
		\multicolumn{1}{|c|}{}                                           & AWidth   
		& 5063.3 & 6477.5 & 5382.0 & 5310.4  \\ \cline{2-6}
		\multicolumn{1}{|c|}{}                                           & FWS   		
		& 8740.9  & 9438.6  & 11394.9 &   9224.4 \\ 
		\hline
		\multicolumn{1}{|c|}{\multirow{1}{*}{Depth method}}  
		& FCov
		& 72.8  & 73.1 & 47.2 & 69.1 \\ \cline{2-6}
		\multicolumn{1}{|c|}{}                                           & AWidth   
		& 5402.4 & 6953.5 & 5935.1 & 5699.9 \\ \cline{2-6}
		\multicolumn{1}{|c|}{}                                           & FWS   		
		& 8914.3 & 10473.1  & 13074.3  & 9738.2 \\ 
		\hline
	\end{tabular}
	\caption{Functional coverage (in $\%$), width and functional Winkler score of the prediction regions for the daily electricity demand based on the SFPL model using the $L_{\infty}$-method, the $\lambda$-method and the depth-based method for each kind of day in 2012  with $\alpha = 0.20$ and $\alpha = 0.05$.}
	\label{tab:PR_demand_SFPL_week_sat_sun}
\end{table}

\begin{figure}[H]
	\centering
	\includegraphics[width=0.4\linewidth]{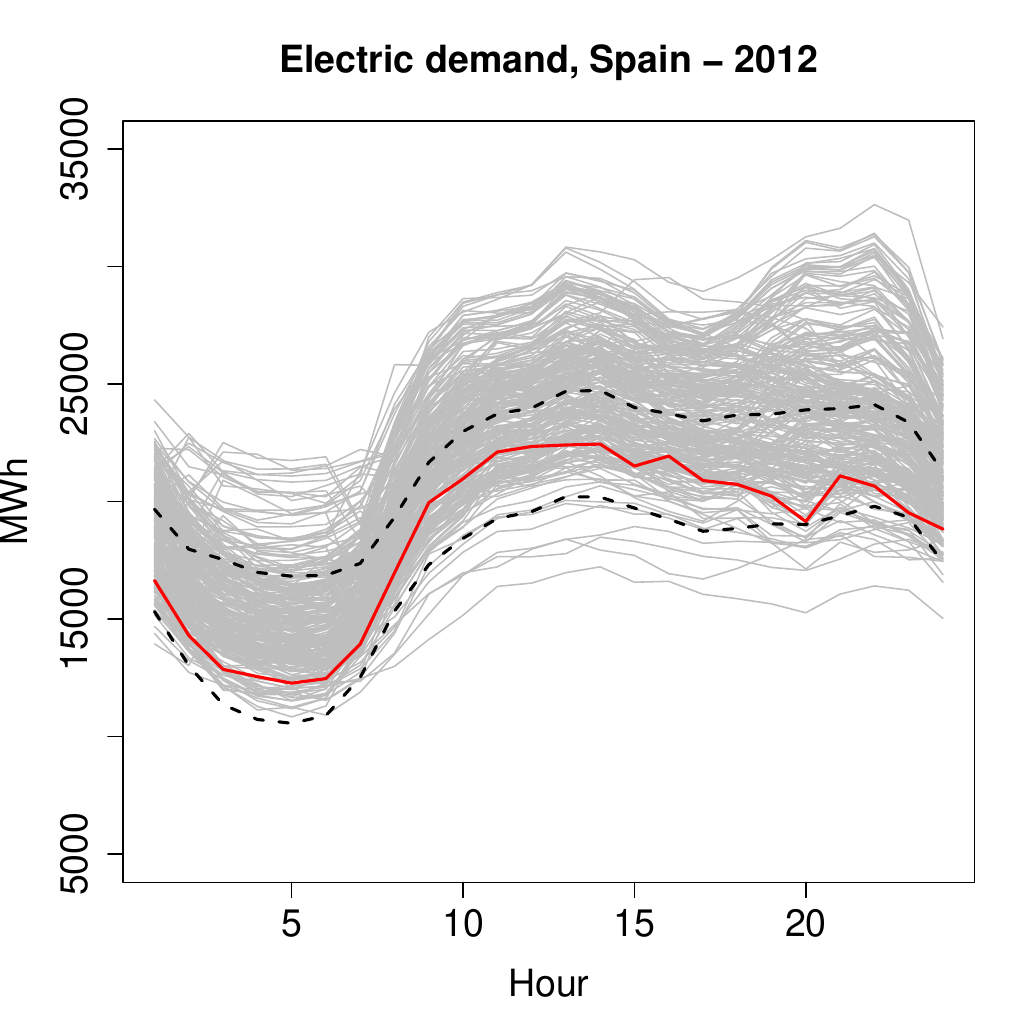}
	\includegraphics[width=0.4\linewidth]{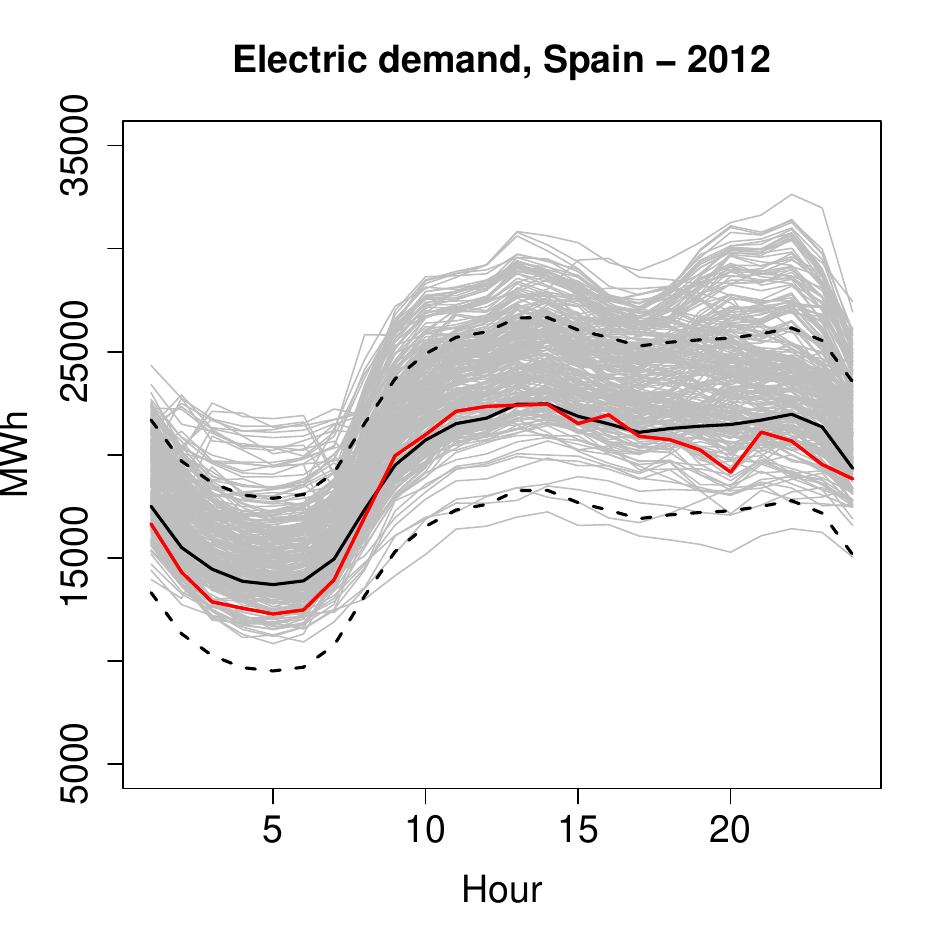}
	\caption{Daily curves of electricity demand corresponding to weekdays (grey solid lines), curve of electricity demand for 2nd April 2012 (red solid line), prediction curve (black solid line) and prediction region (black dashed lines) by means of $\lambda$-method at $80\%$ of confidence (left panel) and $L_{\infty}$-method at $95\%$ of confidence  (right panel) based on the SFPL forecasting model.}
	\label{fig:PR_demand_SFPL_weekdays}
\end{figure}

Figure \ref{fig:PR_demand_SFPL_weekdays} shows the forecast and prediction region of electricity demand on a weekday using the $\lambda$-method at the $80\%$ confidence level and the $L_{\infty}$-method at the $95\%$ confidence level based on the SFPL model. The prediction region obtained at $80\%$ is more informative, as it has a smaller width. The $L_{\infty}$-method provides a region of larger width because the $95\%$ confidence level. Furthermore, the $\lambda$-method takes into account the volatility of electricity demand throughout the day, meanwhile $L_{\infty}$-method provides a region of larger width.

\begin{figure}[H]
	\centering
	\includegraphics[width=0.4\linewidth]{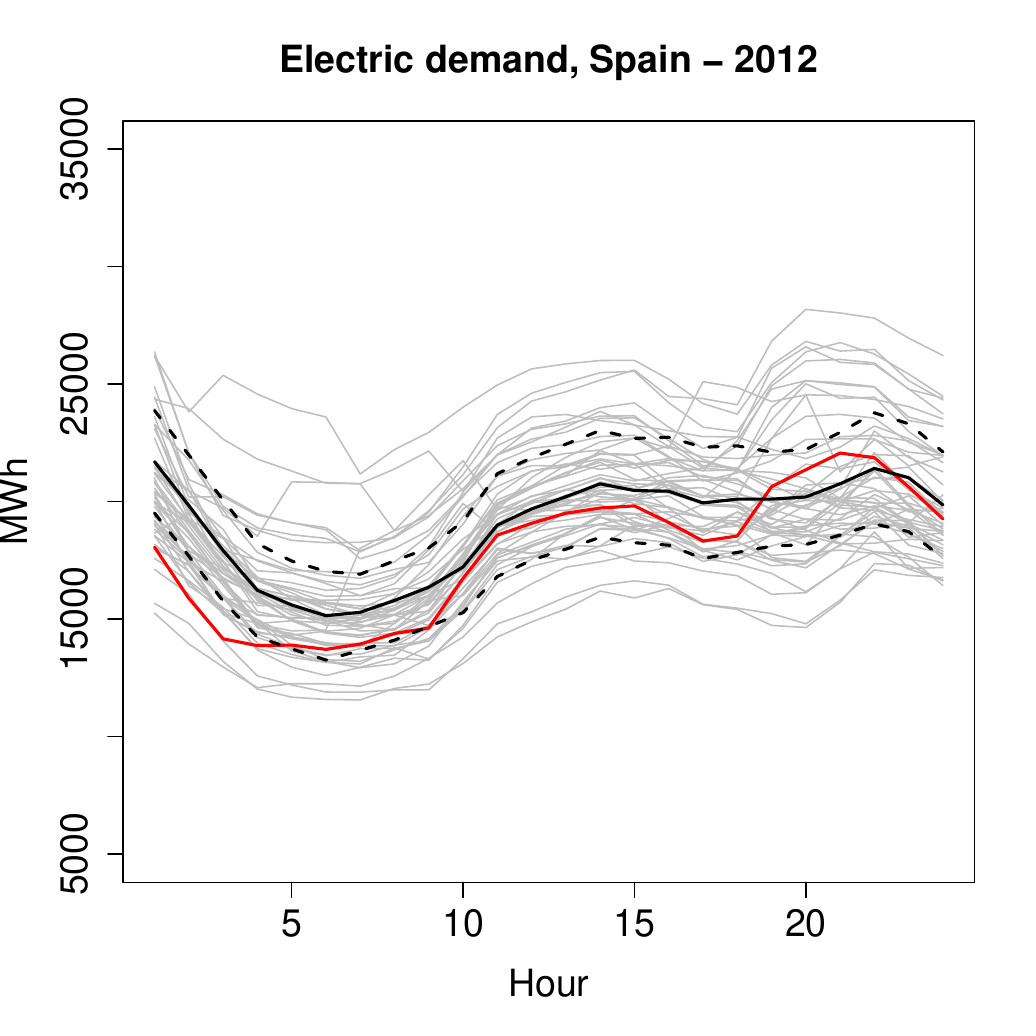}
	\includegraphics[width=0.4\linewidth]{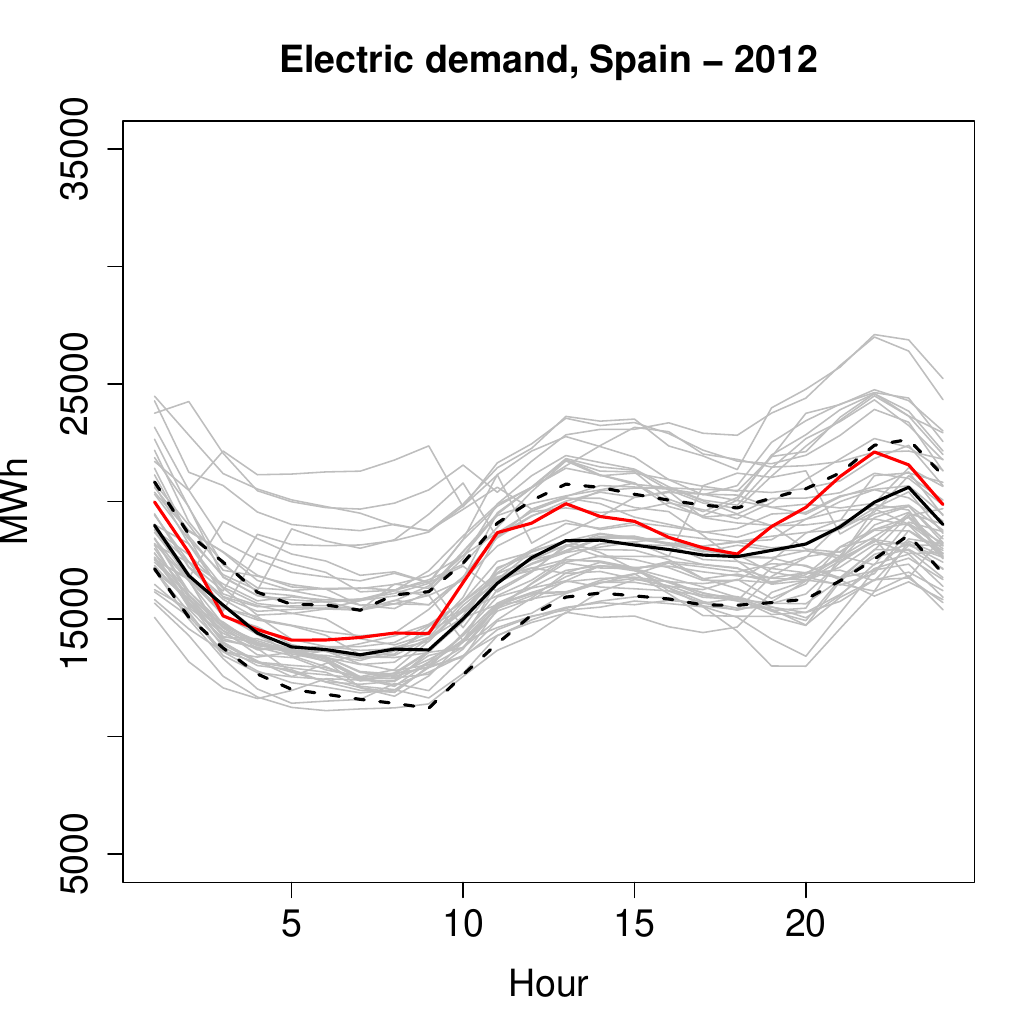}
	\caption{Daily curves of electricity demand corresponding to Saturdays (grey solid lines), curve of electricity demand for 7th January 2012 (red solid line), prediction curve (black solid line) and prediction region (black dashed lines) by means of $\lambda$-method based on the SFPL forecasting model at $80\%$ of confidence (left panel) and daily curves of electricity demand corresponding to Sundays (grey solid lines), curve of electricity demand for 8th January 2012 (red solid line), prediction curve (black solid line) and prediction region (black dashed lines) by means of $\lambda$-method based on the SFPL forecasting model at $80\%$ of confidence (right panel).}
	\label{fig:PR_demand_SFPL_sat_sun}
\end{figure}

Figure \ref{fig:PR_demand_SFPL_sat_sun} shows the forecast and prediction region of electricity demand on a Saturday and a Sunday using the $\lambda$-method based on the SFPL model at the $80\%$ confidence level.
The shortcomings of forecasting on such days are obvious. With less data available and higher volatility, as illustrated by the cloud of curves in Figure \ref{fig:PR_demand_SFPL_sat_sun}, these bootstrap methods perform worse and the resulting prediction regions do not always contain the true daily demand curve.

\subsection{Prediction regions for the electricity price daily curves}
\label{sec:Application_price}

In this section, an analogous study is carried out for the electricity price data.  Prediction regions for the electricity price daily curves are computed using the $L_{\infty}$-method, $\lambda$-method and depth-based method and the forecasting models FNP and SFPL.

Table \ref{tab:PR_price_FNP_week_sat_sun} shows the functional coverage, width and functional Winkler score of the prediction regions obtained with the proposed methods, and using the FNP forecasting model, for the 2012 electricity price curves.  This table compares the different methods for obtaining prediction regions based on the FNP, also taking into account the fixed confidence level. At $95\%$ of confidence, the $L_{\infty}$-method and $\lambda$-method have acceptable functional coverage. The width of the regions obtained with the $\lambda$-method is smaller, which results in a lower FWS. Therefore, the $\lambda$-method seems to be the best alternative in this scenario.
The depth-method gives very small prediction region widths, thus decreasing their coverage. Therefore, it is the one that provides the worst FWS.
The smaller sample size and the higher variability of the curves cause the results for Sundays to be worse.
Fixed the confidence level at $80\%$, the three methods give similar results in this scenario.

\begin{table}[H]
	\centering
	\begin{tabular}{cc|c|c|c|c|}
		\hline
		\multicolumn{2}{|c|}{$\alpha = 0.05$}  & Weekdays & Saturdays   & Sundays & Year \\ \hline
		\multicolumn{1}{|c|}{\multirow{1}{*}{$L_{\infty}$-method}} 
		& FCov
		& 95.4  & 90.4  & 86.8 &  93.4  \\ \cline{2-6}
		\multicolumn{1}{|c|}{}                                           & AWidth   
		& 59.8 & 54.4 & 63.9  & 59.6   \\ \cline{2-6}
		\multicolumn{1}{|c|}{}                                           & FWS   	
		& 92.4  &  105.1  & 145.9  & 102.0  \\ 
		\hline
		\multicolumn{1}{|c|}{\multirow{1}{*}{$\lambda$-method}}  
		& FCov
		& 93.5  & 86.5  &  90.6 & 92.1 \\ \cline{2-6}
		\multicolumn{1}{|c|}{}                                           & AWidth   
		& 47.7  & 47.5 & 58.4   &   49.2 \\ \cline{2-6}
		\multicolumn{1}{|c|}{}         & FWS   	
		& 81.2  & 114.7  & 110.5  & 90.2  \\ \cline{2-6}
		\hline
		\multicolumn{1}{|c|}{\multirow{1}{*}{Depth method}}  
		& FCov
		& 85.1  & 76.9  &  75.5 & 82.5   \\ \cline{2-6}
		\multicolumn{1}{|c|}{}                                           & AWidth   
		& 41.3 & 38.5  &  48.5 & 41.9   \\ \cline{2-6}
		\multicolumn{1}{|c|}{}                                           & FWS   
		& 104.3  & 133.2  & 172.9 & 118.3   \\
		\hline
		\multicolumn{2}{|c|}{$\alpha = 0.20$}  & Weekdays & Saturdays   & Sundays & Year \\ \hline
		\multicolumn{1}{|c|}{\multirow{1}{*}{$L_{\infty}$-method}} 
		& FCov
		& 78.2 & 76.9  & 64.2 &   76.0  \\ \cline{2-6}
		\multicolumn{1}{|c|}{}                                           & AWidth   
		& 33.6 & 33.7  & 38.9  &   34.4 \\ \cline{2-6}
		\multicolumn{1}{|c|}{}                                           & FWS   	
		& 54.8  & 55.5  &  74.6  & 57.8 \\ 
		\hline
		\multicolumn{1}{|c|}{\multirow{1}{*}{$\lambda$-method}}  
		& FCov
		& 68.2 & 73.1  & 60.4  &  67.8  \\ \cline{2-6}
		\multicolumn{1}{|c|}{}                                           & AWidth   
		& 29.9 &  31.2 & 38.4  &   31.4   \\ \cline{2-6}
		\multicolumn{1}{|c|}{}         & FWS   	
		& 55.4  & 54.7 &  75.5 & 58.2   \\ \cline{2-6}
		\hline
		\multicolumn{1}{|c|}{\multirow{1}{*}{Depth method}}  
		& FCov
		&  70.2 &  67.3 &  62.2 & 68.6 \\ \cline{2-6}
		\multicolumn{1}{|c|}{}                                           & AWidth   
		& 30.7 & 31.6  &  40.9 & 32.3 \\ \cline{2-6}
		\multicolumn{1}{|c|}{}                                           & FWS   
		& 54.0  & 60.3  & 82.1  &  59.0  \\ \cline{2-6}
		\hline
	\end{tabular}
	\caption{Functional coverage (in $\%$), width and functional Winkler score of the prediction regions for the daily electricity price based on the FNP model using the $L_{\infty}$-method, the $\lambda$-method and the depth-based method with $\alpha = 0.05$ for each kind of day in 2012.}
	\label{tab:PR_price_FNP_week_sat_sun}
\end{table}

\begin{figure}[H]
	\centering
	\includegraphics[width=0.4\linewidth]{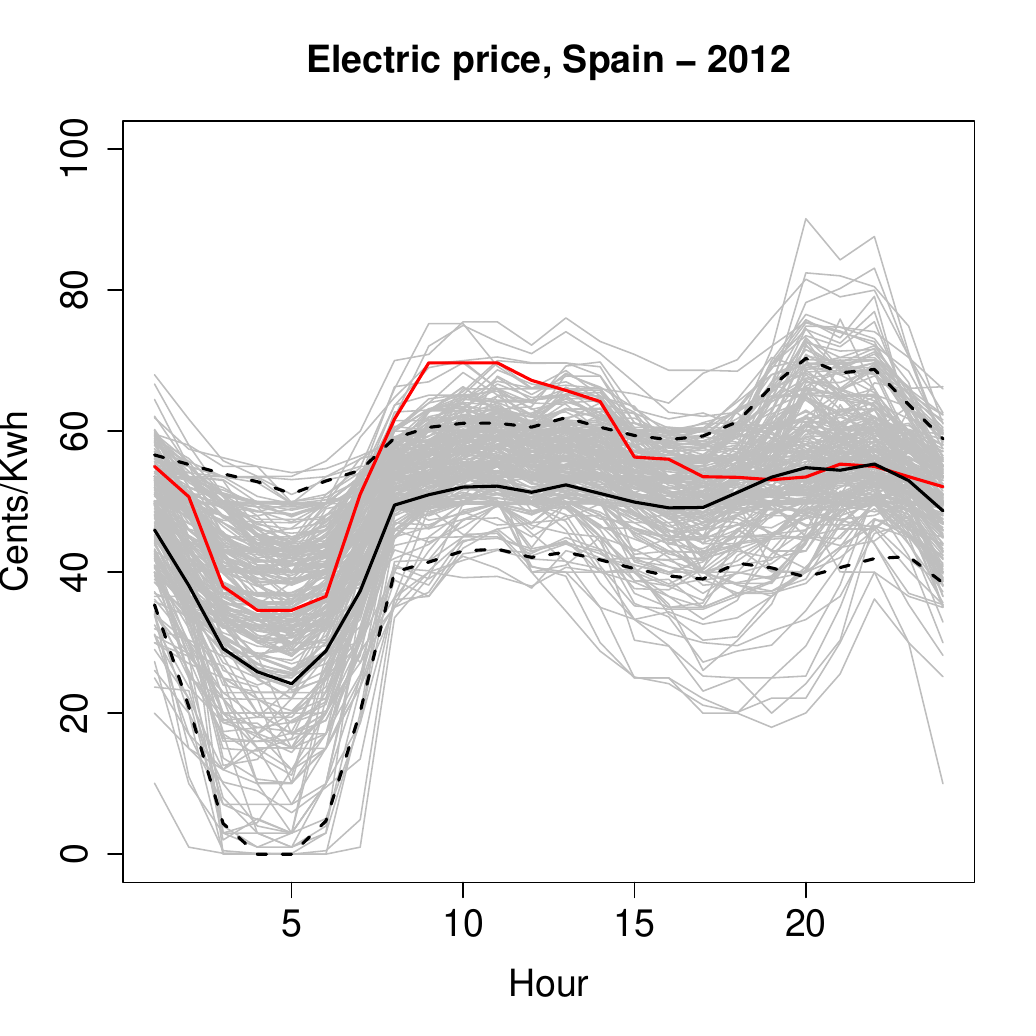}
	\includegraphics[width=0.4\linewidth]{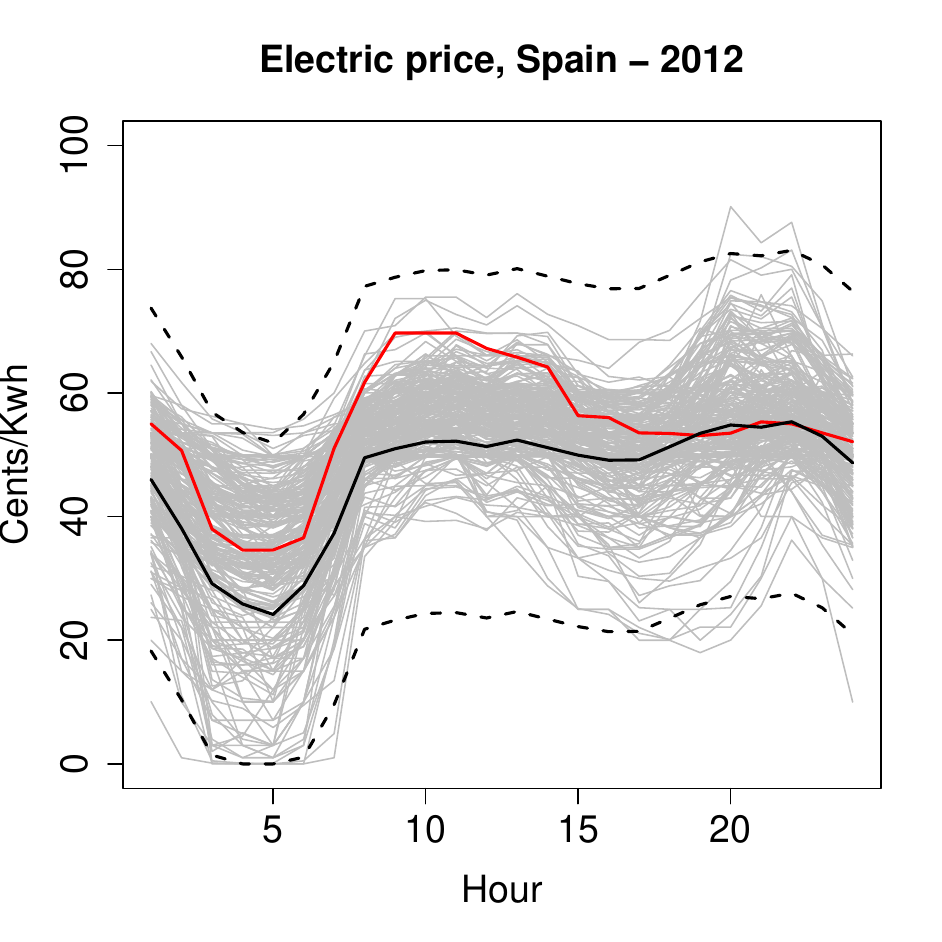}
	\caption{Daily curves of electricity price corresponding to weekdays (grey solid lines), curve of electricity demand for 2nd April 2012 (red solid line), prediction curve (black solid line) and prediction region (black dashed lines) by means of $\lambda$-method at $80\%$ of confidence (left panel) and $L_{\infty}$-method at $95\%$ of confidence  (right panel) based on the FNP forecasting model.}
	\label{fig:PR_price_FNP_weekdays}
\end{figure}

Figure \ref{fig:PR_price_FNP_weekdays} shows the forecast and prediction region of electricity price on a weekday using the $\lambda$-method at the $80\%$ confidence level and the $L_{\infty}$-method at the $95\%$ confidence level based on the FNP model. 
The volatility of the daily electricity price curves is considerable, as can be seen in Figure \ref{fig:PR_price_FNP_weekdays}. This makes the performance of the forecasting models and the methods for computing prediction regions worse. At a confidence level of $95\%$, the methods yield prediction regions of large width and, therefore, not very informative. This is emphasised in the case of the $L_{\infty}$-method when providing prediction regions of constant width.

The methods for obtaining prediction regions for the electricity price daily curves based on the SFPL regression model are discussed below. 
The vector of scalar covariates included in the SFPL model to forecast the daily curves of electricity price is $\mathbf{x}=(x_1,x_2)^T=(\text{D}, \text{PP})^T$ where D is the forecasted daily demand and PP is the wind power production. Other covariates that have a linear relationship with the response variable can be incorporated into the model.

Table \ref{tab:PR_price_SFPL_week_sat_sun} shows the functional coverage, width and Winkler score of the prediction regions obtained with the proposed methods, and using the SFPL forecasting model, for the 2012 electricity demand curves at confidence level $80\%$ and $95\%$ confidence levels. 

The main conclusions drawn above for the prediction region methods based on the FNP model can be extrapolated to the methods based on the SFPL model, although the differences between the $L_{\infty}$-method and the $\lambda$-method based on FNP model are attenuated when using SFPL model. The results for Saturdays and Sundays are still worse than for weekdays.
According to the FWS values obtained, it seems that the results are better with the SFPL model than with the FNP model, especially at $80\%$ confidence.
The performance of the depth method based on the SFPL model is worse than the other two methods, although it is less worse than when using the FNP model.

\begin{table}[H]
	\centering
	\begin{tabular}{cc|c|c|c|c|}
		\hline
		\multicolumn{2}{|c|}{$\alpha = 0.05$}  & Weekdays & Saturdays   & Sundays & Year \\ \hline
		\multicolumn{1}{|c|}{\multirow{1}{*}{$L_{\infty}$-method}} 
		& Coverage
		& 93.5 & 84.6  & 69.8  & 88.8  \\ \cline{2-6}
		\multicolumn{1}{|c|}{}                                           & Width   
		& 49.5 & 40.6  & 41.9 &  47.1 \\ \cline{2-6}
		\multicolumn{1}{|c|}{}                                           & FWS   	
		& 88.8 & 101.8  & 174.4 & 103.1    \\ 
		\hline
		\multicolumn{1}{|c|}{\multirow{1}{*}{$\lambda$-method}}  
		& Coverage
		& 88.1 & 84.6  & 60.4 & 83.6   \\ \cline{2-6}
		\multicolumn{1}{|c|}{}                                           & Width   
		& 40.3 &  34.2 & 37.5 &  39.1  \\ \cline{2-6}
		\multicolumn{1}{|c|}{}         & FWS   	
		& 93.2 & 83.7  & 190.7  &   105.9     \\ \cline{2-6}
		\hline
		\multicolumn{1}{|c|}{\multirow{1}{*}{Depth method}}  
		& Coverage
		& 78.9 & 61.5  & 45.3 &  71.6   \\ \cline{2-6}
		\multicolumn{1}{|c|}{}                                           & Width   
		&  34.2 & 29.5  & 34.5 &  33.6    \\ \cline{2-6}
		\multicolumn{1}{|c|}{}                                           & FWS   
		& 105.9 & 138.8  & 232.2 &  128.9   \\
		\hline
		\multicolumn{2}{|c|}{$\alpha = 0.20$}  & Weekdays & Saturdays   & Sundays & Year \\ \hline
		\multicolumn{1}{|c|}{\multirow{1}{*}{$L_{\infty}$-method}} 
		& Coverage  
		& 73.2  &  63.5 & 45.3 &   67.8  \\ \cline{2-6}
		
		\multicolumn{1}{|c|}{}                                           & Width   
		& 27.9  & 26.5  & 29.2 &    27.9  \\ \cline{2-6}
		\multicolumn{1}{|c|}{}                                           & FWS   	
		& 49.1 & 52.6  & 68.7 & 52.5   \\ 
		\hline
		\multicolumn{1}{|c|}{\multirow{1}{*}{$\lambda$-method}}  
		& Coverage
		& 63.6 &  59.6 & 39.6 &   59.6  \\ \cline{2-6}
		\multicolumn{1}{|c|}{}                                           & Width   
		& 25.6 &  23.2 & 26.9 &  25.4     \\ \cline{2-6}
		\multicolumn{1}{|c|}{}         & FWS   	
		& 50.6 & 45.3  & 69.3 & 52.6 \\ \cline{2-6}
		\hline
		\multicolumn{1}{|c|}{\multirow{1}{*}{Depth method}}  
		& Coverage
		& 59.0 & 50.000  & 39.6 & 54.9   \\ \cline{2-6}
		\multicolumn{1}{|c|}{}                                           & Width   
		& 25.8 &  24.5 & 28.5 &  26.0  \\ \cline{2-6}
		\multicolumn{1}{|c|}{}                                           & FWS   
		& 54.3  & 54.3  & 74.9 &  57.3  \\ \cline{2-6}
		\hline
	\end{tabular}
	\caption{Functional coverage (in $\%$), width and functional Winkler score of the prediction regions for the daily electricity price based on the SFPL model using the $L_{\infty}$-method, the $\lambda$-method and the P-based method with $\alpha = 0.05$ for each kind of day in 2012.}
	\label{tab:PR_price_SFPL_week_sat_sun}
\end{table}

\begin{figure}[H]
	\centering
	\includegraphics[width=0.4\linewidth]{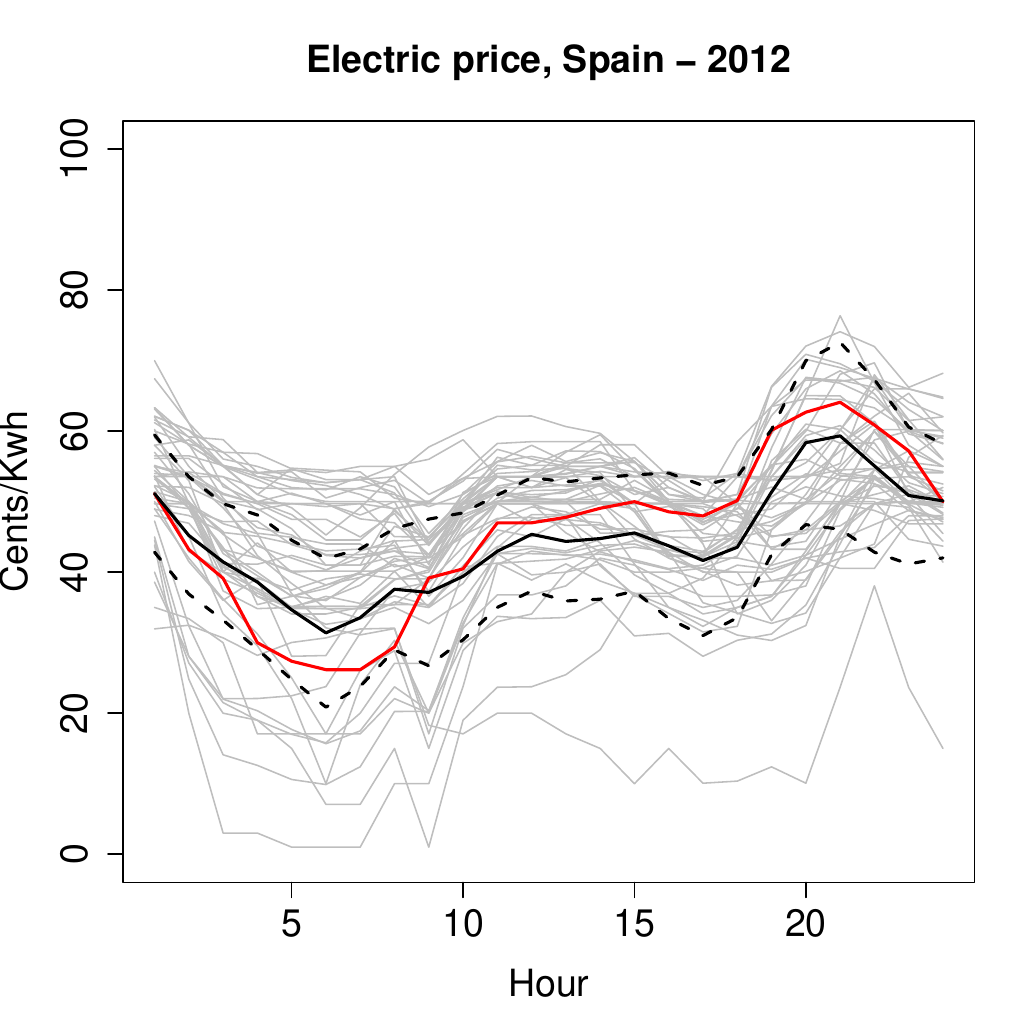}
	\includegraphics[width=0.4\linewidth]{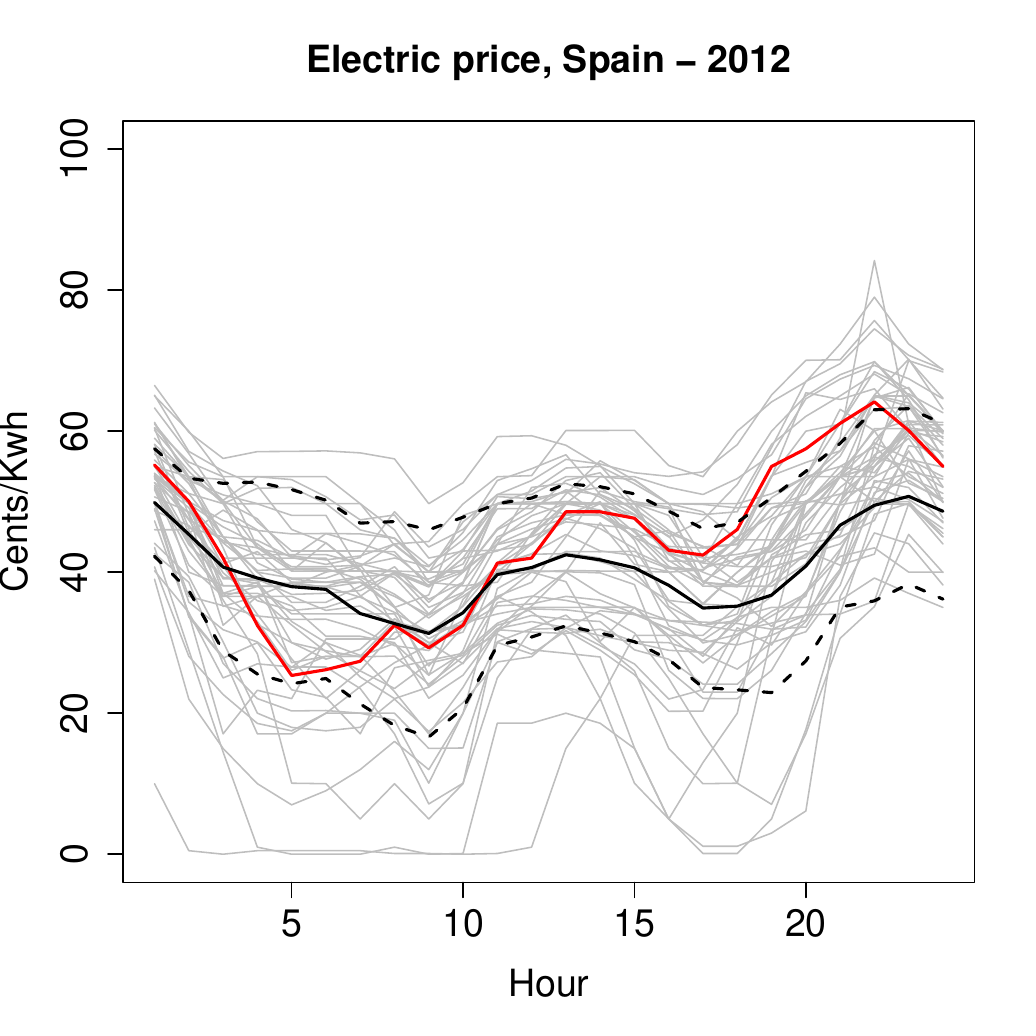}
	\caption{Daily curves of electricity price corresponding to Saturdays (grey solid lines), curve of electricity demand for 7th January 2012 (red solid line), prediction curve (black solid line) and prediction region (black dashed lines) by means of $\lambda$-method based on the SFPL forecasting model at $80\%$ of confidence (left panel) and daily curves of electricity demand corresponding to Sundays (grey solid lines), curve of electricity demand for 8th January 2012 (red solid line), prediction curve (black solid line) and prediction region (black dashed lines) by means of $\lambda$-method based on the SFPL forecasting model at $80\%$ of confidence (right panel).}
	\label{fig:PR_price_SFPL_sat_sun}
\end{figure}

Figure \ref{fig:PR_price_SFPL_sat_sun} shows the forecast and prediction region of electricity price on a Saturday and a Sunday using the $\lambda$-method based on the SFPL model at the $80\%$ confidence level.
As illustrated by Figure \ref{fig:PR_demand_SFPL_sat_sun}, the greater volatility of the daily electricity price curves is increased on Saturdays and Sundays and the bootstrap methods perform worse.

\subsection{Computation time}

An important aspect of these techniques to take into account is their computational cost. 
Table \ref{tab:CPU_time} contains the computational time required to obtain the prediction region for the electricity price or demand curve over a day with each of the three bootstrap methods, $L_{\infty}$-method, $\lambda$-method and depth-based method, based on the two considered regression models, FNP and SFPL. 
The procedures for deriving the functional prediction regions do not differ significantly in terms of computational time. The time differences in practice lie in the regression model chosen.
Although the SFPL model allows more information from the data to be included in the estimation through exogeneous scalar covariates and, therefore, provides more accurate results in some of the analysed scenarios, it has the disadvantage of being computationally more expensive than the FNP model. Specifically, obtaining the one-day-ahead prediction of a daily electricity price or demand curve using the FNP model requires less time than using the SFPL model. Thus, the methods for obtaining FNP model-based prediction regions are also faster than the SFPL model-based methods. This results in large differences between bootstrap methods based on one or the other regression model, as shown in Table \ref{tab:CPU_time}.

\begin{table}[H]
	\centering
	\begin{tabular}{c|r|r|r|}
		\cline{2-4}
		& $L_{\infty}$-method & $\lambda$-method  & depth method  \\ \hline
		\multicolumn{1}{|c|}{FNP}  & 1.8          &  1.8       & 2.0     \\ \hline
		\multicolumn{1}{|c|}{SFPL} & 145.2        &   132.0   &  160.5     \\ \hline
	\end{tabular}
	\caption{Computation time (in seconds) of $L_{\infty}$-method, $\lambda$-method and depth-based method for the computation of the prediction region for one day using FNP and SFPL models.}
	\label{tab:CPU_time}
\end{table}

\section{Conclusions}
\label{sec:Conclusions}

The objective of this paper is to solve the functional regression problem where the response is the daily curve of electricity demand or price, in order to obtain one-day-ahead prediction regions for these curves. To construct the prediction regions, three bootstrap procedures, $L_{\infty}$-method, $\lambda$-method and depth-based method, are proposed. They are based on assuming some functional response forecasting model and bootstraping the residuals. To this extent, these methods are very general, since any functional response regression model could be assumed and any estimator could be used to fit the assumed model. Furthermore, they are able to capture two sources of variability, the one due to the 
estimation of the regression function and the one due to the model error.
In this work, these three general methods are particularised to a functional autoregressive model that is nonparametrically estimated by the Nadaraya-Watson functional estimator (FNP) and a partial linear model with functional response that includes an autoregressive part and linearly introduces scalar covariates (SFPL).

The three proposed techniques have been used to compute prediction regions for the daily curves of both electricity demand and price in the Spanish market. Predictions have been obtained for all days of the year 2012 using information from the previous 365 days. Due to the differences between the types of day, weekday, Saturday and Sunday, it was necessary to fit three different models to predict the electricity demand or price for each day.

The database used in this paper has been analysed in other works on the electricity market. In particular, in \cite{Vilar2018}, they solved the problem of computing pointwise prediction intervals and their results are compared with those of this paper.  The prediction regions of the daily curves obtained from the results of the pointwise (hourly) prediction intervals are very narrow and they have a very low coverage, clearly much lower than the one assumed. Therefore, that technique is discarded and direct procedures based on models with functional response, as the three proposals of this paper, are needed.

The results obtained in the study allow us to conclude that, in general, the $L_{\infty}$-method and the $\lambda$-method are reasonable options for computing prediction regions for the electricity demand and price curves in Spain. Both have fairly similar behaviour, with minor differences. It is important to highlight that the $\lambda$-method takes into account the volatility of the curves, giving rise to prediction regions of non-constant width, an advantage that this method has over the $L_{\infty}$-method.
The depth-based method performed consistently worse than the other two throughout the study.
In general, these observations remain consistent when using both the FNP regression model and the SFPL model. If anything, it is noticed that the use of the SFPL model attenuates the differences between the methods for computing prediction regions.

The proposed procedures for deriving the functional prediction regions do not differ significantly in terms of computational time. The time differences in practice lie in the chosen regression model.
Although the SFPL model allows more information from the data to be included in the estimation through exogeneous scalar covariates and, therefore, provides more accurate results in some of the analysed scenarios, it has the disadvantage of being computationally much more expensive than the FNP model.  It should also be noted that regression models with functional response where an estimation of the daily curve is directly obtained are faster than regression models with scalar response that have to be fitted 24 times, once for each hour or point in the time grid considered. This is another advantage of the techniques presented here, in addition to the improvements in coverage mentioned above.

\section*{Acknowledgements}
This research has been supported by MICINN Grant PID2020-113578RB-100, by the Xunta de Galicia (Grupo de Referencia Competitiva ED431C-2020-14 and Centro Singular de Investigaci\'{o}n de Galicia ED431G 2019/01), all of them through the ERDF.

\bibliographystyle{elsarticle-harv} 
\bibliography{references}

\end{document}